%% file: neurips_2026.tex
\documentclass{article}

 \usepackage[eandd, final]{neurips_2026}

\usepackage[utf8]{inputenc} 
\usepackage[T1]{fontenc}    
\usepackage[hidelinks]{hyperref}       
\usepackage{url}            
\usepackage{booktabs}       
\usepackage{amsfonts}       
\usepackage{nicefrac}       
\usepackage{microtype}      
\usepackage{xcolor}         

\definecolor{gen}{RGB}{88, 146, 66}      
\definecolor{debug}{RGB}{59, 108, 176}   
\definecolor{viz}{RGB}{230, 132, 54}     

\usepackage{amssymb}
\usepackage{bbm}
\usepackage{pifont}
\usepackage{amsmath}
\usepackage{multirow}
\usepackage{makecell}
\usepackage{tikz}
\usepackage{graphicx}
\usepackage{colortbl}
\usepackage{caption}
\usepackage{subcaption}
\usepackage[most]{tcolorbox}
\usepackage{enumitem}
\usepackage{float}

\newcommand{\grayline}{\rowcolor[gray]{.90}}

\title{\textsc{SpreadsheetBench 2}: Evaluating Agents on End-to-End Business Spreadsheet Workflows}

\author{%
  Jian Zhu$^{1,2,3*}$ \quad
  Yuzheng Zhang$^{1,2,3*}$ \quad
  Zeyao Ma$^{1,2}$ \quad
  Bohan Zhang$^{1,2}$ \\
  \textbf{%
  Armin Schoepf$^{4}$ \quad
  Daniel Woloch$^{4}$ \quad
  Peter Yiliu Wang$^{5}$ \quad
  Guangyu Robert Yang$^{5}$} \\
  \textbf{%
  Samuel Jacob$^{6}$ \quad
  Siddharth Nagisetty$^{6}$ \quad
  Abhiram Chundru$^{6}$ \quad
  Jean Lin$^{6}$} \\
  \textbf{%
  Spencer Mateega$^{6}$ \quad
  Jing Zhang$^{1,2\dagger}$%
  } \\
  $^{1}$Renmin University of China, Beijing 100872, China \\
  $^{2}$Key Laboratory of Data Engineering and Knowledge Engineering, MOE, China \\
  $^{3}$Zhongguancun Academy, Beijing, China \\
  $^{4}$Aptura AI
  \quad
  $^{5}$Shortcut AI
  \quad
  $^{6}$AfterQuery \\
  \texttt{jzhu4237@gmail.com} \quad \texttt{zhang-jing@ruc.edu.cn} \\
  $^{*}$Equal contribution. \quad $^{\dagger}$Corresponding author. \\
  \url{https://spreadsheetbench.github.io/}
}

\begin{document}

\maketitle

\begin{abstract}
Spreadsheets are widely used for business analysis, financial modeling, reporting, and decision-making. However, most existing spreadsheet benchmarks evaluate isolated operations such as single-formula generation or local cell edits, and therefore fail to capture end-to-end workflows in realistic business settings. We introduce \textsc{SpreadsheetBench 2}, a workflow-level benchmark for spreadsheet agents that covers three task categories: generation, debugging, and visualization. The benchmark is constructed from authentic business data, including financial reports and corporate filings, and is annotated and validated by domain experts.
The benchmark contains 321 tasks; each instance averages 11.8 worksheets and requires 593.5 cell modifications, reflecting large multi-sheet workbooks with cross-sheet dependencies. We evaluate eight frontier large language models under a unified multi-turn agent scaffold, and additionally include several LLM-based spreadsheet products as complementary baselines. Results show that current systems remain far from reliable on real-world workflows: the best model achieves 34.89\% overall task accuracy, and debugging accuracy is as low as 12.00\%. Trajectory analysis and a failure taxonomy further indicate that insufficient spreadsheet inspection and incorrect target-cell selection are the dominant bottlenecks. Together, these findings position \textsc{SpreadsheetBench 2} as a challenging testbed for advancing reliable spreadsheet automation.
\end{abstract}

\section{Introduction}
\label{sec:introduction}
\input{sections/0introduction}

\section{\textsc{SpreadsheetBench 2}}
\label{sec:spreadsheetbench v2}

\input{sections/1bench_v2}

\section{Experiments}
\label{sec:experiments}
\input{sections/2experiments}

\section{Related Work}
\label{sec:related_work}
\input{sections/3related_work}

\section{Conclusion}
\label{sec:conclusion}
\input{sections/4conclusion}

\begin{ack}
The authors would like to thank the annotators from Aptura AI, Shortcut AI, and AfterQuery who participated in the research and data collection process.
This work is supported by the Zhongguancun Academy, (Grant No.s C20250603).
This work is supported by the National Key Research \& Development Plan (2023YFF0725100) and the National Natural Science Foundation of China (92570121, 62322214, U23A20299, U24B20144).
\end{ack}

\bibliographystyle{unsrt}
\bibliography{ref}

\newpage
\appendix
\input{sections/5appendix}


\newpage
\input{sections/checklist}

\end{document}

%% file: sections/0introduction.tex
Spreadsheets serve as a core infrastructure for structured data processing in modern organizations, supporting tasks such as financial modeling, reporting, and decision-making~\cite{lu2025large, cheng2025survey, broman2018data, taylor2020research, al2022systematic, grossman2005spreadsheet, bradbard2014spreadsheet}. 
Recent advances in large language models (LLMs) have enabled the development of spreadsheet agents capable of generating formulas, manipulating data, and automating a wide range of spreadsheet operations~\cite{li2023sheetcopilot, chen2025sheetagent, wang2026sheetbrain, zhu2025sheetmind, li2024table, joshi2024flame}.
However, despite strong performance on existing benchmarks, these agents remain far from reliable in real-world spreadsheet scenarios.

A key limitation lies in current evaluation benchmarks, which primarily assess spreadsheet agents at the level of isolated operations rather than complete workflows~\cite{li2023sheetcopilot, payan2023instructexcel, ma2024spreadsheetbench}. Existing work often focuses on small or simplified workbooks with limited cross-sheet dependencies~\cite{thorne2025large, iyyer2017search, herzig2020tapas, liu2021tapex}, emphasizing local cell-level manipulations instead of end-to-end task completion. In addition, many datasets are derived from synthetic sources or community forums, resulting in tasks that are less representative of real-world, professional business spreadsheet workflows~\cite{ma2024spreadsheetbench, wu2025tablebench, li2025mimotable, xing2025mmtu, li2023can}.

In practice, spreadsheet usage follows a structured workflow. Tasks commonly involve constructing or completing spreadsheet artifacts, identifying and correcting errors, and presenting results through visualization and reporting. 
These stages introduce challenges such as cross-sheet reasoning, long-range dependencies, and multi-step coordination, which are largely absent from existing benchmarks.

\begin{figure}[t]
\centering
\includegraphics[width=0.9\linewidth]{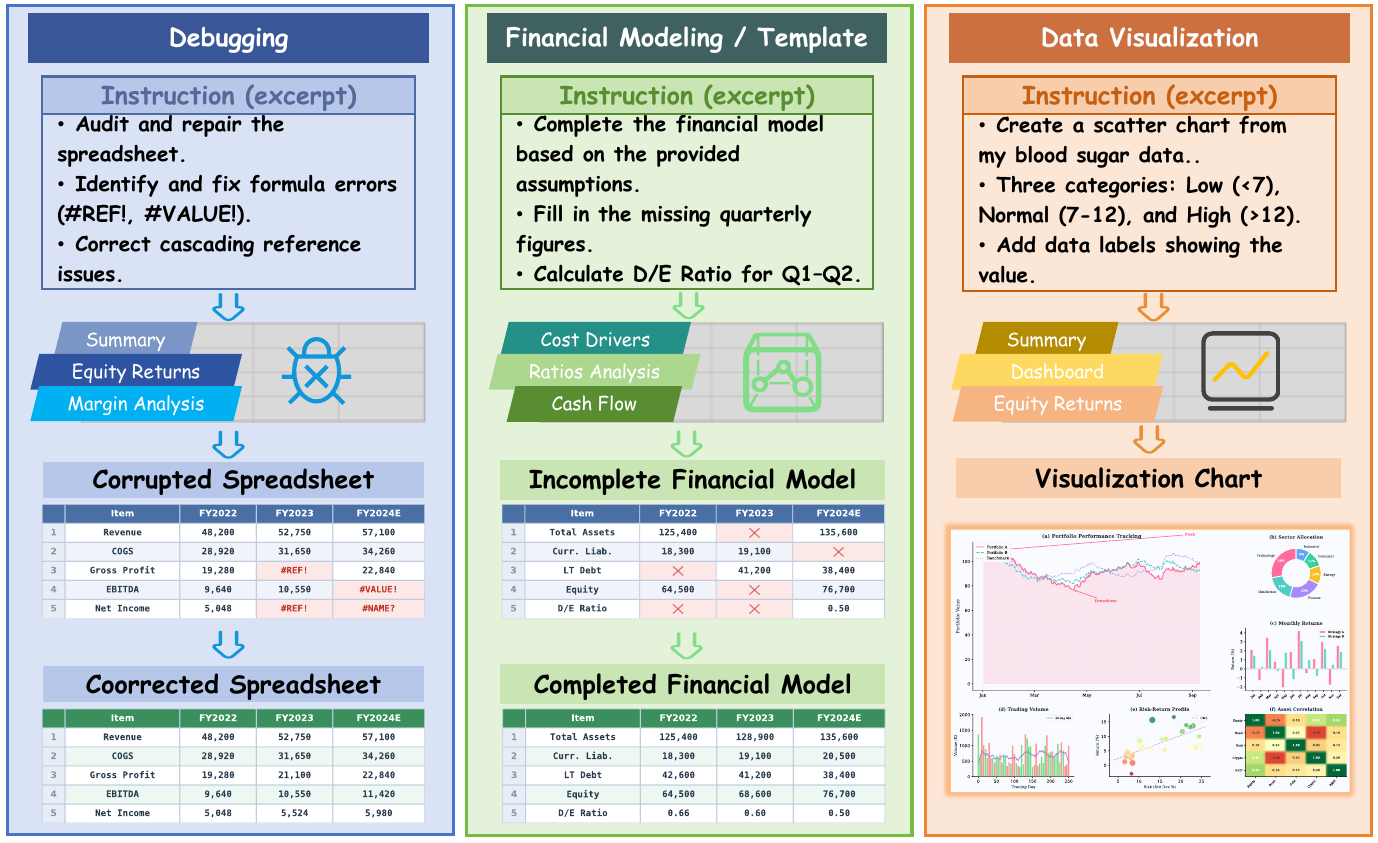}
\caption{
\textsc{SpreadsheetBench 2} consists of three representative task categories: Debugging, Generation (Financial Modeling and Template), and Data Visualization. \textcolor{debug}{Debugging tasks} focus on identifying and repairing errors; \textcolor{gen}{Generation tasks (Financial Modeling and Template)} involve completing or constructing spreadsheets; \textcolor{viz}{Visualization tasks} require producing analysis-ready charts. Detailed examples of each category are provided in Appendix~\ref{app:data examples}.
}
\label{fig:intro}
\end{figure}

To better reflect these characteristics, we introduce \textsc{SpreadsheetBench 2}, a challenging benchmark for evaluating spreadsheet agents on workflow-level tasks grounded in real-world, professional business scenarios. As illustrated in Figure~\ref{fig:intro}, the benchmark organizes tasks into three representative categories: generation (financial modeling and template completion), debugging, and visualization, each corresponding to a key stage in spreadsheet workflows.

\textsc{SpreadsheetBench 2} is constructed from authentic business data sources, including financial reports and corporate filings, and paired with expert-annotated instructions, requiring over 1,500 hours of expert effort (Appendix~\ref{app:annotation effort}). The benchmark consists of 321 tasks, each defined by a natural language instruction and an input spreadsheet representing an in-progress business artifact. These tasks involve complex multi-sheet structures, rich cross-sheet dependencies, and substantial modification requirements, making them significantly more challenging than prior datasets.

We evaluate eight state-of-the-art large language models under a unified multi-turn agent scaffold, where models interact with spreadsheets through a command-line interface. In addition, we include several LLM-based spreadsheet products (e.g., Claude for Excel, ChatGPT for Excel) as complementary baselines. Despite strong performance on existing benchmarks, current LLMs and spreadsheet products achieve limited success on \textsc{SpreadsheetBench 2}, with overall accuracy remaining below 35\% and substantially lower performance on debugging tasks. These results highlight a significant gap between current model capabilities and the requirements of real-world spreadsheet workflows. Further analysis reveals that failures are primarily driven by insufficient inspection and incorrect selection of target cells for modification.
Moreover, to isolate the impact of scaffold design, we fix GLM-5~\cite{zeng2026glm} and find that our SWE-agent~\cite{yang2024swe} based scaffold outperforms three coding agent scaffolds on 50 samples. While modification scores are comparable across scaffolds, task-level accuracy is substantially lower for general-purpose coding scaffolds, indicating weaker end-to-end correctness on spreadsheet workflows.

Our contributions are summarized as follows: (1) We introduce \textsc{SpreadsheetBench 2}, a workflow-level benchmark grounded in real-world business settings, featuring expert-annotated tasks over complex multi-sheet spreadsheets with cross-sheet dependencies, thereby shifting evaluation from isolated operations to end-to-end task completion. (2) We provide a systematic evaluation of state-of-the-art large language models under a unified multi-turn agent scaffold, together with complementary evaluation of LLM-based spreadsheet products, and we further benchmark different agent scaffolds to quantify how scaffold design affects end-to-end performance, revealing substantial performance gaps and key failure modes in real-world spreadsheet automation.

%% file: sections/1bench_v2.tex
\textsc{SpreadsheetBench 2} is designed to evaluate spreadsheet agents on realistic, workflow-level tasks in business settings. 
To achieve this goal, the benchmark is constructed around three key aspects: (1) task formulation based on real-world spreadsheet workflows, (2) data construction grounded in authentic business data, and (3) evaluation protocols that capture both cell-level correctness and task-level outcomes. 
In this section, we describe task categories, dataset construction, and evaluation metrics.

\subsection{Task Categories}

Real-world spreadsheet usage follows a structured workflow rather than isolated operations. 
In business settings, tasks typically involve three stages: constructing spreadsheet artifacts, validating and debugging logical dependencies, and communicating insights through visualization and reporting. 
These stages define the lifecycle of spreadsheet usage in practice. 
Motivated by this workflow perspective, \textsc{SpreadsheetBench 2} organizes tasks into three categories: Generation, Debugging, and Visualization.

\textbf{Generation (Financial Modeling and Template).}
Generation tasks evaluate the ability to construct or complete spreadsheet artifacts from partially specified inputs or high-level objectives. 
These tasks involve cross-sheet reasoning, formula propagation, and the construction of structured spreadsheet. 
They correspond to the construction phase of spreadsheet workflows, where users design and populate structured workbooks from scratch or incomplete specifications. 
We include two levels of difficulty: Template tasks involve simpler, typically single-structure completion (avg. 1.2 sheets), while financial modeling tasks represent more complex scenarios with multiple interdependent components across sheets (avg. 15.3 sheets), requiring coordinated reasoning over large spreadsheet structures.

\textbf{Debugging.}
Debugging tasks evaluate the ability to identify and repair errors in existing spreadsheets. 
In real-world settings, the causes and locations of errors are often unknown. Therefore, instructions do not explicitly specify error types or target cells, requiring models to autonomously locate and fix issues. 
These tasks cover ten error types, including double counting (values are unintentionally aggregated multiple times, leading to systematic overestimation), incorrect references (misconfigured cell links or lookup ranges that break dependencies), formula errors (syntactic or logical mistakes that produce invalid or unintended outputs), and others (Appendix~\ref{app:debugging types}). 
They correspond to the validation and correction phase of spreadsheet workflows and are particularly challenging, as they require holistic reasoning over complex, interdependent spreadsheet structures without explicit information about error locations or types.

\textbf{Visualization.}
Visualization tasks evaluate the ability to transform spreadsheet data into analysis-ready visual artifacts, such as charts and pivot-based summaries. 
In business contexts, spreadsheets function not only as computational tools but also as communication instruments for analysis and decision-making. 
Unlike standard visualization benchmarks, the input data in these tasks is often irregularly structured, distributed across multiple sheets, or embedded within semi-structured tables, requiring models to correctly interpret and extract relevant information. 
The target visualizations can involve complex structures, such as multiple data series, hierarchical axes, and customized formatting. 
These tasks correspond to the presentation stage of spreadsheet workflows and require both accurate data selection and precise visual construction within spreadsheet-native formats.

\begin{figure*}[t]
\centering
\includegraphics[width=1\linewidth]{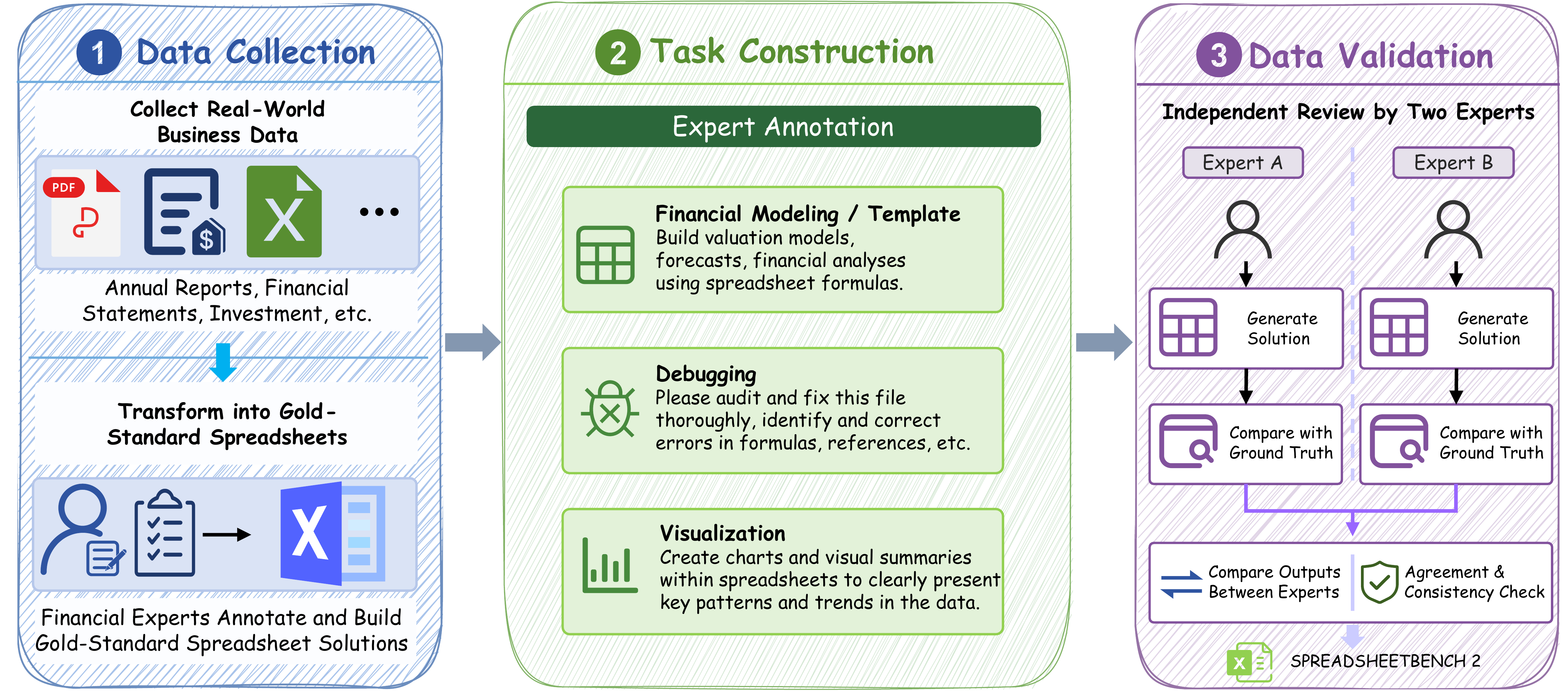}
\caption{
The benchmark construction pipeline of \textsc{SpreadsheetBench 2}.
}
\label{fig:data_pipeline}
\end{figure*}

\subsection{Benchmark Construction}
As shown in Figure~\ref{fig:data_pipeline}, \textsc{SpreadsheetBench 2} is constructed through a three-stage pipeline: data collection, task construction, and data validation.

\textbf{Data Collection.}
We construct \textsc{SpreadsheetBench 2} using data from authentic business sources, including publicly available financial reports and corporate filings from the \textit{NYU Stern (Damodaran) dataset}, \textit{Screener.in}, \textit{Bloomberg}, \textit{Bseindia} (More details in Appendix~\ref{app:data sources}).
Financial experts curate these materials and build complete, internally consistent spreadsheets as gold-standard solutions, with correct formulas, cross-sheet references, and domain-appropriate logic.
To capture the diversity of real-world spreadsheet workflows, the dataset covers a wide range of financial and accounting scenarios, including accounting tasks (e.g., consolidation, deferred tax, equity investments) and finance-related analyses such as discounted cash flow (DCF), leveraged buyout (LBO), mergers and acquisitions (M\&A), and comparable company analysis.

\textbf{Task Construction.}
Based on the gold-standard spreadsheets, each task is created by transforming a complete workbook into a partially specified input spreadsheet paired with a natural-language instruction.
Depending on the task category, this transformation involves selectively removing target regions, injecting controlled errors, or specifying visualization objectives, such that the resulting task requires the model to recover or produce the missing content.
All tasks are manually designed and annotated by financial experts to reflect realistic business spreadsheet workflows.
The construction process enforces several quality constraints: each task must admit a unique, deterministic solution given only the input spreadsheet and instruction; the instruction must be self-contained without requiring external knowledge beyond standard financial domain conventions; and the task must involve multi-step reasoning over complex spreadsheet structures rather than isolated cell-level edits.
As a result, solving a single task typically requires coordinated operations across multiple sheets, including formula propagation, reference resolution, structural inference, and consistency verification, mirroring the cognitive demands of real-world spreadsheet work in professional settings.

\textbf{Data Validation.}
To ensure the overall quality of the benchmark, each task is independently reviewed by two experts who were not involved in its construction. 
Reviewers are provided only with the input spreadsheet and the natural-language instruction, and are asked to solve the task independently. 
Their solutions are then compared against the original ground-truth spreadsheet. 
Reviewers also inspect both the input and ground-truth workbooks for unintended defects.
If discrepancies arise, the task is iteratively revised through expert discussion and correction until consistent solutions are obtained.
Unresolved disagreements are adjudicated by a senior expert.
This cross-validation process ensures that task instructions are unambiguous, all required operations are well-defined, and the final solution is uniquely determined by the given spreadsheet and instruction. All 321 released tasks satisfied the acceptance criteria within three validation rounds. More details on the validation process and statistics are provided in Appendix~\ref{app:data_validation}.

\input{table/benchmark_stat}

\subsection{Benchmark Statistics}

\textbf{Data Statistics.}
\textsc{SpreadsheetBench 2} comprises 321 tasks across three categories: generation (financial modeling and template), debugging, and visualization. Table~\ref{tab:overall_stats} summarizes the key statistics.
The benchmark comprises an average of 11.8 worksheets and 593.5 modified cells per task, with Financial Modeling and Debugging emerging as the most challenging categories, averaging 1,164.5 and 656.6 modified cells per task, respectively.
Solving a single task thus requires reasoning over cross-worksheet dependencies and coordinating hundreds of cell-level modifications, a level of complexity characteristic of real-world professional spreadsheet use.

\textbf{Comparison with existing benchmarks.}
Table~\ref{tab:benchmark_comparison} compares \textsc{SpreadsheetBench 2} with representative spreadsheet benchmarks across multiple dimensions.
First, \textsc{SpreadsheetBench 2} is grounded in real-world business data curated by domain experts, whereas SheetCopilot~\cite{li2023sheetcopilot} and InstructExcel~\cite{payan2023instructexcel} rely on synthetic instructions and SpreadsheetBench~\cite{ma2024spreadsheetbench} draws from semi-real forum posts.
Second, our tasks exhibit substantially higher structural complexity: an average of 11.8 sheets per file compared to roughly 1--2 in prior work, with instructions averaging 429.0 words that specify multi-step business objectives rather than single atomic operations.
Third, \textsc{SpreadsheetBench 2} introduces task dimensions absent from existing benchmarks, including a systematic error taxonomy for debugging, native spreadsheet chart generation for visualization, and end-to-end workflow-level evaluation that requires sustained cross-sheet reasoning and domain-specific knowledge.
These properties make \textsc{SpreadsheetBench 2} a more realistic and challenging testbed for evaluating agent capabilities in professional spreadsheet environments.

\input{table/compare_exist_bench}

\subsection{Evaluation Metrics}
\label{sec:eval metrics}

We adopt two evaluation protocols tailored to different output types. For Financial Modeling, Template, and Debugging tasks, we compare the agent's output against the golden spreadsheet. For Visualization tasks, whose outputs are charts, we employ a VLM-based evaluation.

\textbf{Spreadsheet Execution Metrics.}
By comparing the input spreadsheet with the golden file, we identify the set of \emph{target cells}---those whose values or formulas must be changed to complete the task.
We report two complementary metrics that operate at different granularities.
\textit{Modification} is a \emph{cell-level} metric: for each task, we compute the fraction of target cells whose computed values match the golden file; we then report the average across all tasks in a category.
It captures how precisely the agent performs the required edits, independent of side effects elsewhere.
\textit{Accuracy} is a \emph{task-level} metric: a task is scored 1 if every cell in the output spreadsheet matches the golden file (covering both target and non-target cells), and 0 otherwise; we report the fraction of tasks that achieve this full match.

\textbf{VLM-as-a-Judge.}
Rule-based verification is impractical for visualization tasks due to the diversity of chart types, layouts, and formatting. Instead, we design fine-grained \textit{rubrics} for each task along two dimensions: \textit{Data Correctness} (e.g., whether plotted values match the source data) and \textit{Format Compliance} (e.g., chart type, axis labels, color schemes). All rubrics are constructed and validated by domain experts (Appendix~\ref{app:rubric example}). We use GLM-4.6V~\cite{hong2025glm} as the judge to produce a binary \textsc{pass}/\textsc{fail} signal per criterion, and report the rubric pass rate as the task score.
We further assess evaluator robustness through comparisons across three VLM judges, rubric-level majority voting, and a human audit of the Claude Opus~4.6 outputs. Model rankings remain largely stable across judges, while majority voting achieves 94.5\% overall agreement with human annotations on the audited outputs (Appendix~\ref{app:vlm_judge_robustness}).

%% file: table/benchmark_stat.tex
\begin{table}[t]
\centering
\small
\caption{Overview statistics of \textsc{SpreadsheetBench 2}. Avg. Sheets and Max. Sheets correspond to the mean and maximum number of worksheets per task file, respectively. Avg. Words indicates the length of the input instructions, while Modif. Cells denotes the average number of cells requiring modification.}
\label{tab:overall_stats}
\resizebox{0.85\textwidth}{!}{%
\begin{tabular}{@{}lccccc@{}}
\toprule
\textbf{Category} 
& \textbf{\#Tasks} 
& \textbf{Avg. Sheets} 
& \textbf{Max. Sheets} 
& \textbf{Avg. Words} 
& \textbf{Modif. Cells} \\
\midrule
Financial Modeling & 100 & 15.3 & 43 & 705.9 & 1164.5 \\
Debugging & 100 & 21.1 & 99 & 387.0 & 656.6 \\
Template & 97 & 1.2 & 3 & 139.9 & 65.7 \\
Visualization & 24 & 1.7 & 8 & 617.7 & 88.3 \\
\midrule
\textbf{Overall} & 321 & 11.8 & 99 & 429.0 & 593.5 \\
\bottomrule
\end{tabular}%
}
\end{table}


%% file: table/compare_exist_bench.tex
\newcommand{\cmark}{\textcolor{green!60!black}{\ding{51}}}
\newcommand{\xmark}{\textcolor{red!70!black}{\ding{55}}}

\begin{table}[t]
\centering
\caption{Comparison of \textsc{SpreadsheetBench 2} with existing benchmarks, emphasizing real-world tasks, multi-sheet and cross-sheet reasoning, and high domain expertise.}
\label{tab:benchmark_comparison}
\small
\setlength{\tabcolsep}{3pt}
\resizebox{\textwidth}{!}{%
\begin{tabular}{@{}lcccc@{}}
\toprule
\textbf{Aspect} & \textbf{SheetCopilot~\cite{li2023sheetcopilot}} & \textbf{InstructExcel~\cite{payan2023instructexcel}} & \textbf{SpreadsheetBench~\cite{ma2024spreadsheetbench}} & \textbf{Ours} \\
\midrule
Realism & Synthetic & Synthetic & Semi-real & Real-world + Expert-curated \\
Total Tasks & 221 & 4,850 & 912 & 321 \\
Avg. Instr. Words & 27.9 & 9.8 & 85.7 & 429.0 \\
Multi-Sheet Tasks & Rare & Limited & Limited & Extensive \\
Domain Expertise & Low & Low & Low & High \\
Avg. Sheets/File & $\sim$1 & $\sim$2 & $\sim$1.4 & 11.8 \\
Error Taxonomy & \xmark & \xmark & \xmark & \cmark \\
Chart Tasks & \xmark & \xmark & \xmark & \cmark \\
Real Workflow & \xmark & \xmark & \xmark & \cmark \\
\bottomrule
\end{tabular}%
}
\end{table}

%% file: sections/2experiments.tex
\input{table/main_result}
\subsection{Experimental Settings}

\textbf{LLMs.}
We evaluate eight large language models via API-based inference, including five open-source models (GLM-5.0~\cite{zeng2026glm}, MiniMax M2.5~\footnote{\url{https://www.minimax.io/news/minimax-m25}}, Kimi K2.5~\cite{team2026kimi}, Qwen3.5-397B-A17B~\cite{qwen35blog}, and DeepSeek-V3.2~\cite{liu2025deepseek}) and three closed-source models (GPT-5.2~\footnote{\url{https://openai.com/index/introducing-gpt-5-2/}}, Gemini 3.1 Pro~\footnote{\url{https://blog.google/innovation-and-ai/models-and-research/gemini-models/gemini-3-1-pro/}}, and Claude Opus 4.6~\footnote{\url{https://www.anthropic.com/news/claude-opus-4-6}}). All models are evaluated in thinking mode, with GPT-5.2 and Gemini 3.1 Pro configured to use \textit{high} reasoning effort. In addition, we include several LLM-based spreadsheet tools (Kimi Sheet, GLM in Excel, Claude for Excel, and ChatGPT for Excel) for complementary comparison; as these products do not expose a unified API interface compatible with our evaluation pipeline, we evaluate them through manual execution on a subset of 30 representative examples.

\textbf{Agent Scaffold.}
Our agent scaffold is built on SWE-agent~\cite{yang2024swe} and retains its iterative observe--reason--act workflow~\cite{yao2023react}. In the main setting, the agent interacts with spreadsheets through three tools: \texttt{bash}, \texttt{view\_xlsx}, and \texttt{submit}. The \texttt{bash} tool executes local commands for programmatic workbook operations, \texttt{view\_xlsx} provides read-only inspection of workbook structure and sheet contents (including formulas and evaluated values), and \texttt{submit} finalizes the episode for scoring. We set the maximum number of interaction turns to 50 and keep the scaffold fixed across models by using the same prompt template and tool specifications (additional implementation details are provided in Appendix~\ref{app:details of experiments}). We further compare GLM-5 across different agent scaffolds, including our scaffold, Claude Code, Kilo Code, and Cline, to study the impact of scaffold design.

\subsection{Main Results}

\textbf{\textsc{SpreadsheetBench 2} remains challenging even for frontier models.}
As shown in Table~\ref{tab:main_results}, the best-performing model, Claude Opus 4.6, achieves only 34.89\% overall accuracy, and six of the eight evaluated models fall below 25\%.
Closed-source models hold a clear advantage over their open-source counterparts: the top three closed-source models reach 23.68\%--34.89\% accuracy, substantially outperforming all open-source models (7.17\%--17.14\%).
The gap is equally pronounced within individual categories: Claude Opus~4.6 achieves 89.69\% Modification but only 34.00\% Accuracy on Financial Modeling, and 50.38\% Modification but only 12.00\% Accuracy on Debugging.
This suggests that while current LLMs can perform individual cell-level edits with moderate success, they struggle to maintain cross-cell consistency and end-to-end correctness across multi-step workflows---indicating that the bottleneck lies not in isolated operations but in coordinating coherent modifications over complex, interdependent spreadsheet structures.

\textbf{Performance varies significantly across task categories.}
Table~\ref{tab:main_results} presents the accuracy scores of various models across different task categories. The results reveal a pronounced performance disparity among task types: visualization tasks exhibit the lowest difficulty, with Claude Opus 4.6 achieving the highest accuracy of 62.5\%; generation tasks, specifically template and financial modeling, follow with moderate performance; whereas debugging tasks pose the greatest challenge, with even the best-performing model, Claude Opus 4.6, attaining only 12\% accuracy. These findings indicate that while current large language models have developed strong capabilities in spreadsheet visualization and foundational structure generation, they still face significant performance bottlenecks in real-world business scenarios that demand complex logical reasoning, error diagnosis, and automated correction.

\begin{figure}[t]
\centering
\includegraphics[width=0.95\linewidth]{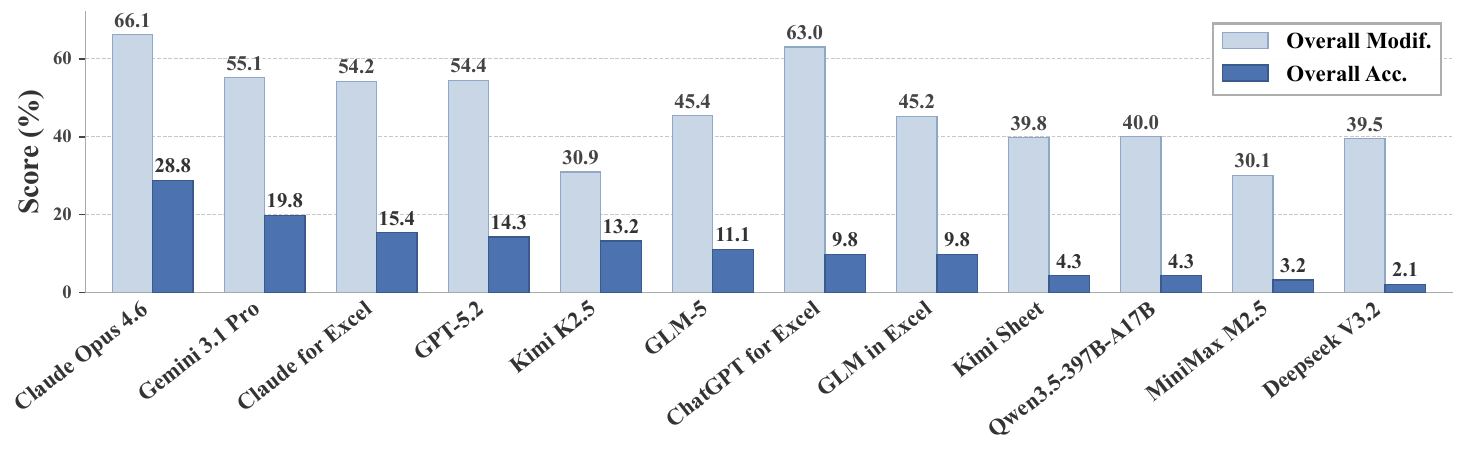}
\caption{
Performance on a 30-example representative subset covering Financial Modeling, Debugging, and Visualization tasks. We compare different models and four LLM-based spreadsheet products. The results show that LLM-based spreadsheet products do not outperform models under our agent scaffold, with Claude for Excel achieving the best performance among the spreadsheet products.
}
\label{fig:subset_score}
\end{figure}

\textbf{LLM-based spreadsheet products do not show clear performance advantages.}
As shown in Figure~\ref{fig:subset_score}, we compare four LLM-based spreadsheet products\textemdash Kimi Sheet, GLM in Excel, Claude for Excel, and ChatGPT for Excel\textemdash against foundation models on a challenging subset of 30 examples spanning Financial Modeling, Debugging, and Visualization (10 each). For a fair comparison, each product is tested by human operators given only the task instruction and input file. The results show that none of the spreadsheet products surpass the foundation models, with Claude for Excel achieving the highest accuracy among them at only 15.4\%, suggesting that the workflow-level complexity of \textsc{SpreadsheetBench 2} poses significant challenges for current LLM-based spreadsheet products.

\subsection{Analysis and Discussions}

\begin{figure}[t]
\centering
\includegraphics[width=0.95\linewidth]{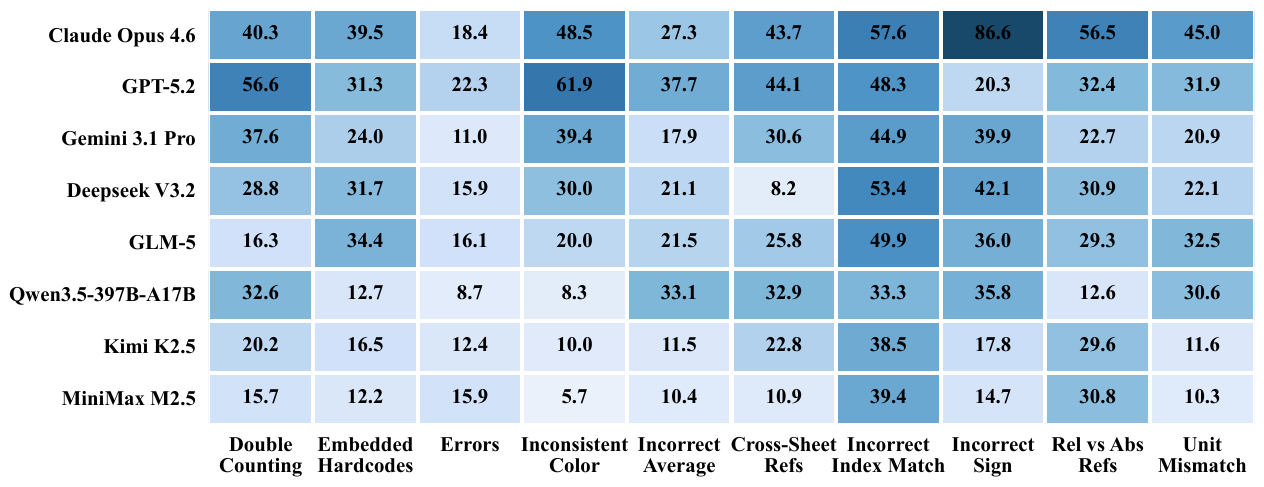}
\caption{
Modification scores across 10 error subcategories (Appendix~\ref{app:debugging types}) in Debugging tasks.
}
\label{fig:debug_heatmap}
\end{figure}

\textbf{Ten Error Types of Debugging Tasks.}
In Figure~\ref{fig:debug_heatmap}, we compare model performance across ten debugging error subcategories (see Table~\ref{tab:debug_taxonomy}). Difficulty varies: \textit{Incorrect Index Match} and \textit{Incorrect Sign} are relatively tractable, while \textit{Errors} is the most challenging, with even the best model barely exceeding 20\%. This category involves diagnosing and fixing formulas that produce explicit spreadsheet error values (e.g., \#REF!, \#NUM!), which often stem from complex cross-sheet dependencies or invalid references, making them particularly difficult to localize and resolve.
Models exhibit complementary strengths, with Claude Opus 4.6 performing best on \textit{Incorrect Sign} and GPT-5.2 leading on \textit{Double Counting} and \textit{Inconsistent Color}, and no single model dominates across all subcategories.
Notably, performance drops markedly on \textit{Errors}, highlighting the persistent difficulty of diagnosing and fixing formula-level failures in spreadsheets.

\begin{figure}[t]
    \centering
    \begin{subfigure}[t]{0.32\linewidth}
        \centering
        \includegraphics[width=\linewidth]{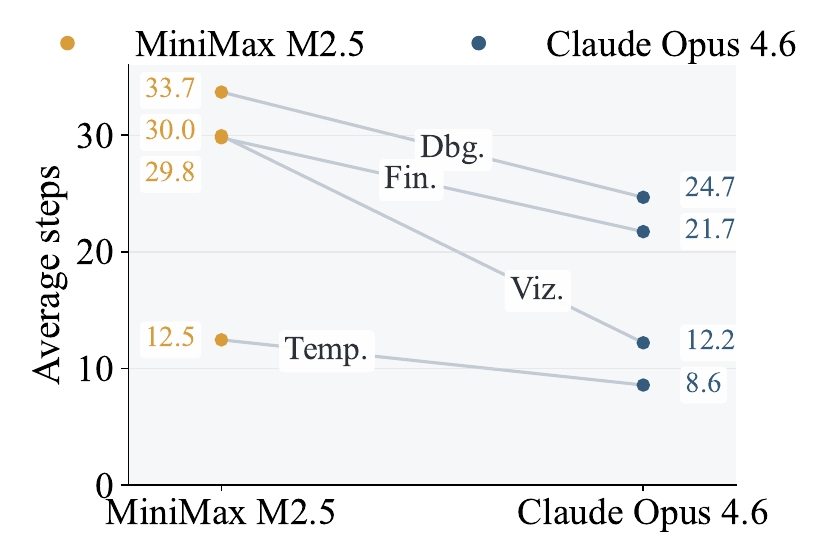}
        \caption{}
        \label{fig:7a}
    \end{subfigure}
    \hfill
    \begin{subfigure}[t]{0.32\linewidth}
        \centering
        \includegraphics[width=\linewidth]{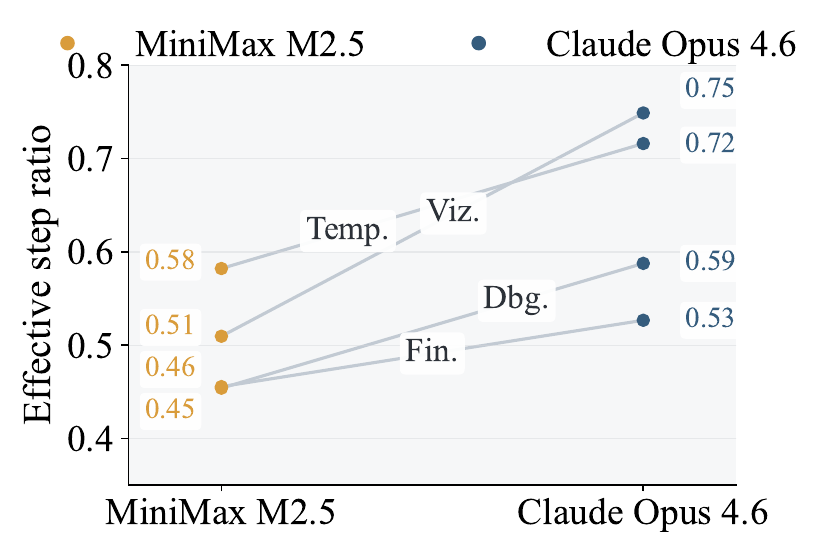}
        \caption{}
        \label{fig:7b}
    \end{subfigure}
    \hfill
    \begin{subfigure}[t]{0.34\linewidth}
        \centering
        \includegraphics[width=\linewidth]{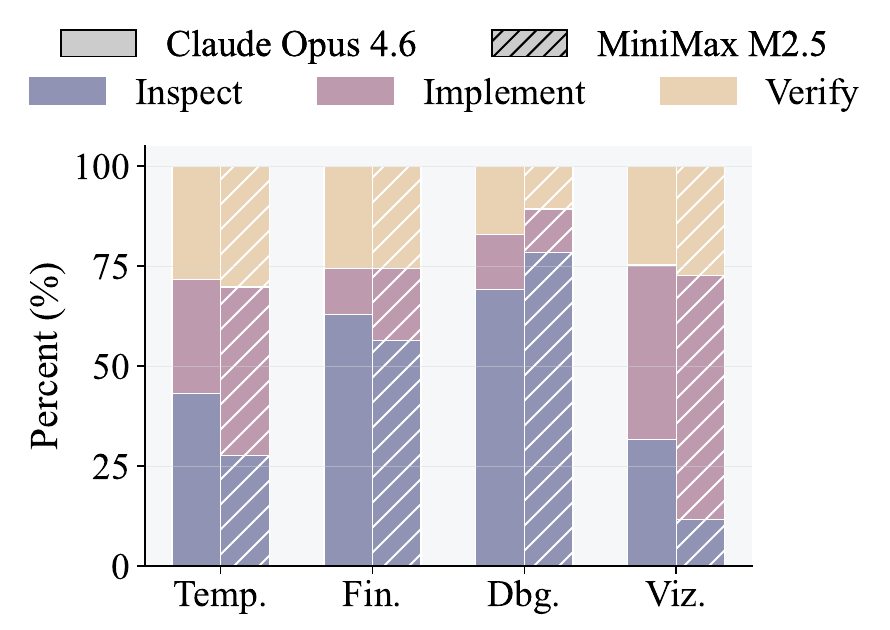}
        \caption{}
        \label{fig:7c}
    \end{subfigure}
    \caption{Overview of trajectory-level behavior across four task domains. 
    (a)~Average number of steps per task for each model. 
    (b)~Effective-step ratio, measuring the fraction of steps that directly contribute to task progress. 
    (c)~Process distribution, showing the proportion of inspect, implement, and verify phases within each domain.}
    \label{fig:7}
\end{figure}

\textbf{Interaction Turns and Effective Step Ratio.}
We compare Claude Opus 4.6 and MiniMax M2.5 in Figure~\ref{fig:7}(\subref{fig:7a}). Claude Opus 4.6 consistently achieves higher modification scores and accuracy across all task categories than MiniMax M2.5. It also requires fewer average interaction steps across four task types, suggesting that increased interaction-step count does not necessarily improve performance in spreadsheet-based tasks. To explain this efficiency gap, we define an \textit{effective step} as one executed without errors (e.g., tool-call failures, code execution errors, or bash failures). Figure~\ref{fig:7}(\subref{fig:7b}) reveals that Claude Opus 4.6 maintains a substantially higher effective-step ratio, confirming that its advantage stems from more reliable per-step execution rather than a larger interaction budget: each step more consistently translates into meaningful task progress.

\noindent
\begin{minipage}[t]{0.65\textwidth}
\textbf{Failure Taxonomy.}
To elucidate the underlying causes of agent failures, we conducted a systematic analysis of failure trajectories, categorizing errors into six distinct types, as summarized in Table~\ref{tab:failure_modes}, where the detailed definitions of each type are provided. Leveraging Claude Code for automated trajectory analysis of Claude Opus 4.6, we derived the error distribution illustrated in Figure~\ref{fig:error_taxonomy_donut}. Our results indicate that the predominant error categories are \textit{Insufficient Inspection} and \textit{Wrong Target Selection}. This suggests that agents struggle to adequately comprehend the intricate interdependencies within spreadsheets—such as formulaic, numerical, and domain-specific logical relationships—and fail to accurately pinpoint the target cells requiring modification. Notably, these challenges reflect common pain points experienced by humans in real-world spreadsheet workflows.
\end{minipage}%
\hfill
\begin{minipage}[t]{0.33\textwidth}
\centering
\captionof{figure}{Failure Taxonomy distribution for Claude Opus 4.6 trajectories of unresolved tasks.}
\label{fig:error_taxonomy_donut}
\includegraphics[width=\linewidth]{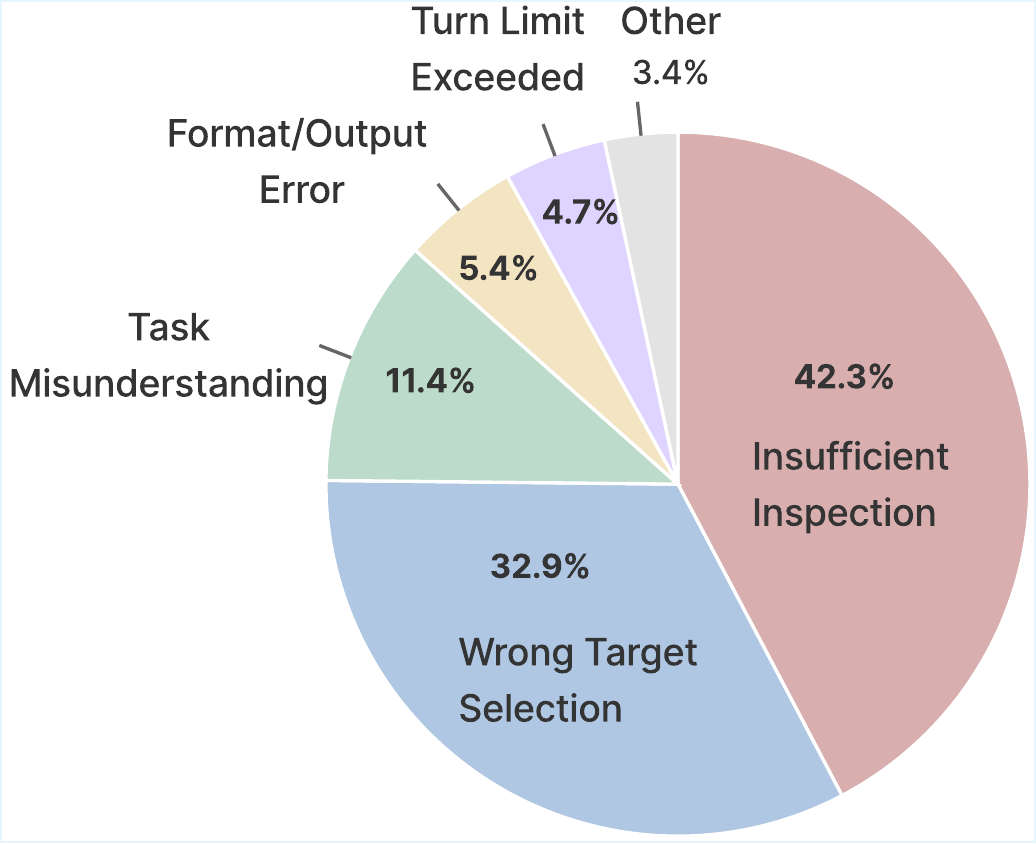}
\end{minipage}



\textbf{Tool Usage Patterns.}
To understand the behavioral patterns of agents in spreadsheet tasks, we categorize tool usage trajectories into three functional phases (Figure~\ref{fig:7}(\subref{fig:7c})): \textit{Inspect} (reading contents and examining formulas), \textit{Implement} (writing cells and executing code), and \textit{Verify} (re-reading modified cells to confirm correctness). Claude Opus 4.6 exhibits a higher inspection proportion than MiniMax M2.5 in Template, Visualization, and Financial Modeling tasks, but this trend reverses in Debugging tasks, where both models allocate over 60\% of their trajectory to inspection. We attribute this to the absence of explicit step-by-step instructions in Debugging tasks, which necessitates extensive sheet exploration and imposes significant demands on the agent’s ability to localize information within complex spreadsheet environments.

\noindent
\begin{minipage}[t]{0.65\textwidth}
\textbf{Different Agent Scaffold.}
Using GLM-5 as the fixed backbone, we compare our SWE-agent-based scaffold against three coding agent scaffolds on a subset of 50 samples (Table~\ref{tab:scaffold_overall}). Our scaffold achieves the highest modification score (46.32\%) and accuracy (15.45\%), with Claude Code as the closest competitor (43.47\% / 14.20\%). 
While modification scores remain within a narrow 4-point band, accuracy diverges sharply: Cline and Kilo Code attain only 8.66\%, suggesting that general-purpose coding scaffolds can produce plausible edits but struggle with end-to-end correctness on spreadsheet tasks.
\end{minipage}%
\hfill
\begin{minipage}[t]{0.33\textwidth}
\centering
\captionof{table}{Overall performance comparison across different agent scaffolds.}
\label{tab:scaffold_overall}
\resizebox{\linewidth}{!}{%
\begin{tabular}{lcc}
\toprule
Agent Scaffold & Modif. & Acc. \\
\midrule
Our Scaffold & \textbf{46.32} & \textbf{15.45} \\
Claude Code & 43.47 & 14.20 \\
Cline & 41.98 & 8.66 \\
Kilo & 43.12 & 8.66 \\
\bottomrule
\end{tabular}%
}
\end{minipage}

\textbf{Case Study.}
We highlight a representative Financial Modeling failure (details in Appendix~\ref{app:case_study}, Figure~\ref{fig:case4}).
The task requires filling six groups of empty cells across a multi-sheet banking model.
Claude Opus~4.6 derived formulas from general financial knowledge (e.g., ``Book Value per Share = Total Equity / Shares Outstanding'') rather than reverse-engineering the conventions already established in the workbook; the formulas are textbook-correct but the cell references did not match the actual row layout, yielding only 4.1\% modification accuracy.
This reveals that current agents can retrieve relevant domain knowledge but lack the ability to ground it in a specific workbook's dependency structure, a capability gap that future spreadsheet agents must address.
Five additional cases spanning Debugging and Visualization are provided in Appendix~\ref{app:case_study}.

%% file: table/main_result.tex
\definecolor{pbgray}{RGB}{240, 240, 240}
\definecolor{modifcol}{RGB}{144, 195, 244}
\definecolor{acccol}{RGB}{255, 194, 112}
\definecolor{bestrow}{RGB}{232, 243, 255}

\newcommand{\pbar}[2]{%
  \begin{tikzpicture}[baseline=(orig.base), inner sep=0pt, outer sep=0pt]
    \def\barwidth{1.1cm}
    \def\barheight{0.32cm}
    \fill[pbgray, rounded corners=1pt] (0,0) rectangle (\barwidth, \barheight);
    \pgfmathsetmacro{\fw}{(#1/100)*1.1}
    \fill[#2, rounded corners=1pt] (0,0) rectangle (\fw cm, \barheight);
    \node (orig) at (0.55cm, 0.16cm) {\fontsize{8}{8}\selectfont #1};
  \end{tikzpicture}%
}

\newcommand{\pbarb}[2]{%
  \begin{tikzpicture}[baseline=(orig.base), inner sep=0pt, outer sep=0pt]
    \def\barwidth{1.1cm}
    \def\barheight{0.32cm}
    \fill[pbgray, rounded corners=1pt] (0,0) rectangle (\barwidth, \barheight);
    \pgfmathsetmacro{\fw}{(#1/100)*1.1}
    \fill[#2, rounded corners=1pt] (0,0) rectangle (\fw cm, \barheight);
    \node (orig) at (0.55cm, 0.16cm) {\fontsize{8}{8}\selectfont \textbf{#1}};
  \end{tikzpicture}%
}

\newcommand{\pbm}[1]{\pbar{#1}{modifcol}}
\newcommand{\pba}[1]{\pbar{#1}{acccol}}
\newcommand{\pbmb}[1]{\pbarb{#1}{modifcol}}
\newcommand{\pbab}[1]{\pbarb{#1}{acccol}}

\begin{table}[t]
\centering
\small
\caption{Performance of open-source and closed-source models across four spreadsheet task categories. \textit{Modif.} reports the average fraction of target cells whose computed values match the golden file. \textit{Acc.} reports the fraction of tasks where \emph{every} cell in the output matches the golden file. For Visualization tasks, Acc.\ instead reports the average rubric pass rate (fraction of expert-designed rubric satisfied per task). The Overall column aggregates across all applicable tasks per metric.}
\label{tab:main_results}
\setlength{\tabcolsep}{4pt}
\renewcommand{\arraystretch}{1.5}
\resizebox{\linewidth}{!}{
\begin{tabular}{lcccccccccc}
\toprule
\multirow{2}{*}{Model}
  & \multicolumn{2}{c}{Template}
  & \multicolumn{2}{c}{Financial Modeling}
  & \multicolumn{2}{c}{Debugging}
  & \multicolumn{1}{c}{Visualization}
  & \multicolumn{2}{c}{Overall} \\
\cmidrule(lr){2-3}\cmidrule(lr){4-5}\cmidrule(lr){6-7}
\cmidrule(lr){8-8}\cmidrule(lr){9-10}
  & Modif. & Acc.
  & Modif. & Acc.
  & Modif. & Acc.
  & Acc.
  & Modif. & Acc. \\
\midrule
\grayline
\multicolumn{10}{c}{\textit{Closed-source Models}} \\
\midrule
\rowcolor{bestrow}
\textbf{Claude Opus 4.6}
  & \pbmb{91.97} & \pbab{52.58}
  & \pbmb{89.69} & \pbab{34.00}
  & \pbmb{50.38} & \pbab{12.00}
  & \pbab{62.50}
  & \pbmb{77.20} & \pbab{34.89} \\
GPT-5.2
  & \pbm{83.84} & \pba{35.05}
  & \pbm{87.07} & \pba{33.00}
  & \pbm{39.07} & \pba{8.00}
  & \pba{45.83}
  & \pbm{69.85} & \pba{26.79} \\
Gemini 3.1 Pro
  & \pbm{86.06} & \pba{28.87}
  & \pbm{85.26} & \pba{31.00}
  & \pbm{41.86} & \pba{7.00}
  & \pba{41.67}
  & \pbm{70.91} & \pba{23.68} \\
\midrule
\grayline
\multicolumn{10}{c}{\textit{Open-source Models}} \\
\midrule
GLM-5
  & \pbm{66.60} & \pba{17.53}
  & \pbm{78.44} & \pba{22.00}
  & \pbm{29.06} & \pba{7.00}
  & \pba{37.50}
  & \pbm{57.95} & \pba{17.14} \\
Deepseek-V3.2
  & \pbm{75.19} & \pba{25.77}
  & \pbm{59.78} & \pba{7.00}
  & \pbm{28.71} & \pba{10.00}
  & \pba{33.33}
  & \pbm{54.35} & \pba{15.58} \\
Kimi K2.5
  & \pbm{67.08} & \pba{18.56}
  & \pbm{64.86} & \pba{15.00}
  & \pbm{19.48} & \pba{4.00}
  & \pba{41.67}
  & \pbm{50.31} & \pba{14.64} \\
Qwen3.5-397B-A17B
  & \pbm{65.62} & \pba{17.53}
  & \pbm{69.59} & \pba{10.00}
  & \pbm{24.06} & \pba{3.00}
  & \pba{25.00}
  & \pbm{52.96} & \pba{11.22} \\
MiniMax M2.5
  & \pbm{50.14} & \pba{7.22}
  & \pbm{62.05} & \pba{8.00}
  & \pbm{16.76} & \pba{4.00}
  & \pba{16.67}
  & \pbm{42.91} & \pba{7.17}  \\
\bottomrule
\end{tabular}%
}
\end{table}

%% file: sections/3related_work.tex
\textbf{Spreadsheet and Table Benchmarks.}
A substantial body of work has developed benchmarks for table understanding, spanning question answering over structured tables, fact verification against tabular evidence, and numerical reasoning over hybrid tabular-textual data~\cite{pasupat2015compositional, chen2019tabfact, chen2020hybridqa, zhu2021tat, chen2021finqa, chen2022convfinqa, zhao2022multihiertt, wu2025tablebench}.
On the spreadsheet side, prior efforts address two complementary directions: table encoding and representation learning for spreadsheet structures~\cite{yin2020tabert, dong2024spreadsheetllm}, and benchmarking spreadsheet manipulation from natural language instructions, including formula generation, cell editing, and task-level completion~\cite{li2023sheetcopilot, payan2023instructexcel, zhao2024nl2formula, ma2024spreadsheetbench}.
However, these benchmarks predominantly target isolated atomic operations, such as generating a single formula or editing specific cells, rather than the multi-step, cross-sheet workflows that characterize real business spreadsheet usage.
\textsc{SpreadsheetBench 2} addresses this gap with end-to-end workflow-level tasks spanning generation, debugging, and visualization, averaging 11.8 sheets and 593.5 modified cells per instance.

\textbf{LLM-based Spreadsheet Agents.}
Recent advances in LLM-based agents have demonstrated the potential of autonomous tool use for complex tasks, with agent frameworks and benchmarks emerging across coding, web navigation, and desktop environments~\cite{yao2023react, schick2023toolformer, wang2024executable, jimenez2023swe, zhou2023webarena, liu2023agentbench}.
In the spreadsheet domain, several agent systems have been proposed to automate spreadsheet operations, adopting diverse strategies such as translating natural language into API calls, multi-module reasoning pipelines, neuro-symbolic computation, and multi-agent collaboration~\cite{li2023sheetcopilot, chen2025sheetagent, wang2026sheetbrain, zhu2025sheetmind}.
Despite this progress, existing spreadsheet agents are evaluated only on isolated operations, leaving a gap in assessing performance on realistic business workflows that require sustained cross-sheet reasoning and coordinated tool use.
\textsc{SpreadsheetBench 2} provides a workflow-level evaluation testbed that requires agents to perform multi-step, cross-sheet reasoning and produce structurally coherent spreadsheet artifacts, enabling systematic comparison of agent capabilities on realistic business tasks.

%% file: sections/4conclusion.tex
We present \textsc{SpreadsheetBench 2}, a benchmark of 321 expert-annotated tasks evaluating spreadsheet agents on workflow-level generation, debugging, and visualization over multi-sheet workbooks with rich cross-sheet dependencies. The best model achieves only 34.89\% accuracy (debugging: 12\%), and stronger models benefit from more reliable per-step execution rather than larger interaction budgets, highlighting structured reasoning over interdependent workbooks as a key bottleneck.

%% file: sections/5appendix.tex
\section{Broader Discussion}
\label{app:broader_discussion}

\subsection{Limitations}
\label{app:limitations}

Although \textsc{SpreadsheetBench 2} advances spreadsheet agent evaluation, several limitations remain.
(1) The benchmark focuses on financial and business spreadsheet workflows, and the findings may not generalize to other spreadsheet-intensive domains such as scientific computing or engineering.
(2) Due to the high cost of API-based LLM inference, we do not report error bars or confidence intervals for the experimental results.
(3) For visualization tasks, the VLM-as-a-judge framework may introduce evaluation noise despite expert-designed rubrics; for cell-based tasks, exact-match evaluation may undercount semantically equivalent but syntactically different solutions.

\subsection{Ethical Considerations}
\label{app:ethical}

In this section, we discuss the ethical considerations regarding our data construction process.
(1) Data Risk Control: During the benchmark construction process, we review all spreadsheet files and task instructions to ensure they do not contain content that is inappropriate for a general audience, such as violence, discrimination, or sensitive subjects. In addition, two authors conduct an independent verification to confirm that the benchmark does not contain any personally identifiable information. All financial data is derived from publicly available corporate filings and financial reports.
(2) Copyright: Our benchmark is constructed from publicly available financial reports and corporate filings. To prevent copyright infringement, we do not directly redistribute any raw source documents. Instead, the released benchmark consists of expert-constructed spreadsheet tasks derived from these public materials, with numerical values and structural layouts modified where necessary. Considering potential licensing and copyright risks, we release the benchmark under a CC BY-SA 4.0 license and refrain from secondary distribution of any raw source data.

\subsection{Broader Impacts}
\label{app:impacts}

On the positive side, advancing spreadsheet agent capabilities can reduce manual effort in repetitive business workflows, democratize access to financial modeling, and help detect costly spreadsheet errors that are a well-documented source of financial losses.
On the negative side, over-reliance on AI-based spreadsheet agents without adequate human oversight could lead to undetected errors in financial models, and automation of financial tasks may raise concerns about workforce displacement.
Our benchmark explicitly quantifies the gap between current model capabilities and real-world requirements, with the best model achieving only 34.89\% overall accuracy, discouraging premature deployment and encouraging further research on reliability.

\subsection{License}
\label{app:license}

The \textsc{SpreadsheetBench 2} dataset is released under the CC BY-SA 4.0 license, which permits sharing and adaptation, including commercial use, with appropriate credit and under the same license. The evaluation code is released under the MIT License. All third-party assets used in this work are properly cited and their respective licenses are respected.

\section{Details of Data}

\subsection{Data Sources}
\label{app:data sources}

The financial data underlying \textsc{SpreadsheetBench 2} is drawn from a diverse set of publicly available sources.
These include primary regulatory filings from \textit{Bseindia}~\footnote{\url{https://www.bseindia.com/corporates/ann}}, \textit{Bloomberg}~\footnote{\url{https://www.bloomberg.com/}}, \textit{Screener.in}~\footnote{\url{https://www.screener.in/}}, and \textit{the NYU Stern (Damodaran) dataset}~\footnote{\url{https://pages.stern.nyu.edu/~adamodar/New_Home_Page/data.html}}.
For sources with redistribution restrictions, we release only derived task metadata and evaluation artifacts when permitted, while separately documenting access requirements and licensing constraints.
Together, these sources provide broad coverage of real-world business data across multiple geographies and industry sectors, ensuring the diversity and authenticity of the benchmark tasks.
All source materials are publicly available at the time of collection, and no proprietary or restricted-access data is used in the construction of the benchmark.

\subsection{Annotation Effort}
\label{app:annotation effort}
Our annotation team consisted of 20 finance professionals, each with at least two years of relevant industry experience, serving as senior experts, annotators, or independent reviewers. Senior experts with more than five years of experience defined the taxonomy and adjudicated unresolved cases. Annotators with more than two years of experience created the task instructions, input workbooks, and gold-standard spreadsheets. Reviewers with more than three years of experience independently validated tasks they had not constructed.

The construction process required over 1,500 hours of expert annotation, with approximately 400 hours devoted to Financial Modeling, 500 to Debugging, 380 to Visualization, and 200 to Template tasks. These estimates exclude additional quality assurance rounds conducted internally and by our data provider partners, including having human solvers complete selected tasks to verify their solvability.

\subsection{Data Validation}
\label{app:data_validation}

\textbf{Validation Process.}
Each candidate task was independently reviewed by two finance professionals who had not participated in its annotation. Validation proceeded in two stages. First, reviewers attempted the task using only the natural-language instruction and input workbook, without consulting the gold-standard workbook, to assess solvability, ambiguity, and the required financial reasoning. Second, they inspected both the input and gold-standard workbooks for unintended errors, file corruption, and unexpected spreadsheet behavior.

Each reviewer independently assessed three pass/fail criteria:
\begin{enumerate}[leftmargin=*]
    \item The input and gold-standard workbooks are free of unintended defects, excluding errors deliberately introduced into the input workbooks of Debugging tasks.
    \item The task can be completed using the provided instruction and input workbook without unresolved ambiguity or missing information.
    \item The task requires multi-step reasoning and financial domain knowledge.
\end{enumerate}

Reviewers documented failed criteria and discussed the necessary revisions. Unresolved disagreements were adjudicated by a senior expert, who determined whether the task should be revised or discarded. Tasks selected for revision were returned to the annotators and subsequently re-reviewed until all three criteria were satisfied.

\textbf{Validation Statistics.}
The validation process retained 321 tasks from approximately 400 initial candidates. All released tasks satisfied the acceptance criteria within three validation rounds. Table~\ref{tab:validation_statistics} summarizes the acceptance rounds, task revisions and exclusions, and reviewer consistency.

\input{table/validation_statistics}

\subsection{Visualization Task Rubric Examples}
\label{app:rubric example}
We evaluate visualization tasks using a VLM-as-a-judge approach with hand-crafted rubrics that assess two complementary aspects: \emph{Format Compliance} (chart type, axis configuration, color scheme) and \emph{Data Correctness} (value ranges, data labels). Table~\ref{tab:rubric_example} shows the complete rubric for a football field valuation chart (Task~129), illustrating the granularity of our per-criterion evaluation.
\input{table/rubric_example}

\subsection{Debugging Task Types}
\label{app:debugging types}
Table~\ref{tab:debug_taxonomy} presents the taxonomy of spreadsheet error types used in the Debugging task. These ten categories were identified through iterative annotation of real-world spreadsheet errors encountered in financial modeling and accounting practice. The taxonomy spans a range of failure modes, from low-level syntactic issues such as formula errors and incorrect sign assignments, to higher-level structural problems including cross-sheet reference failures and unit mismatches. Each debugging instance contains one or more injected errors drawn from this taxonomy, and the agent is required to localize and correct all errors to restore the spreadsheet to its intended state.
\input{table/debug_taxonomy}

\section{Details of Experiments}
\label{app:details of experiments}

\subsection{Environment.}
Each task instance is executed inside a sandboxed Docker container built on Python~3.11 (Debian Bullseye). The container is pre-installed with common data-science libraries (NumPy, Pandas, Matplotlib, openpyxl) and LibreOffice with its Python UNO bridge, enabling both programmatic and application-level spreadsheet manipulation. No network access is provided during execution, ensuring that the agent must rely solely on its own reasoning and the provided tools.

\subsection{Tools.}
We conduct our experiments within the SWE-Agent framework. To rigorously evaluate the intrinsic capabilities of the model, we restrict the available toolset to three fundamental utilities: bash, view\_xlsx, and submit.

\begin{itemize}[leftmargin=*]
    \item \textbf{bash}: Executes arbitrary shell commands, providing the agent with general-purpose programmatic access to the operating system. Common usage includes file operations (\texttt{ls}, \texttt{cat}, \texttt{cp}, \texttt{mv}), running Python scripts via \texttt{python3} for data manipulation and formula computation, and invoking standard Unix utilities for text processing (\texttt{grep}, \texttt{awk}, \texttt{sed}).
    \item \textbf{view\_xlsx}: A read-only utility for inspecting \texttt{.xlsx} spreadsheet files. The agent must specify one of two modes: \texttt{list}, which enumerates all sheet names in the workbook, or \texttt{content}, which displays the contents of a specific sheet. The \texttt{content} mode optionally accepts a row range to limit the output (e.g., rows 1--50), and exposes both the original formulas and their evaluated values.
    \item \textbf{submit}: Finalizes the agent's solution and submits the modified spreadsheet for evaluation. The agent is expected to invoke this tool only after all edits have been completed and verified.
\end{itemize}

\subsection{Prompt.}
The agent receives a two-part prompt at each episode. The \emph{system prompt} (Figure~\ref{fig:system_prompt}) establishes the agent's role and interaction protocol, while the \emph{instance prompt} supplies the task-specific context, including available tools, a recommended workflow, and the concrete instructions with input/output file paths. We use two instance prompt variants: one for Template, Financial Modeling, and Debugging tasks (Figure~\ref{fig:instance_prompt}), and a separate one for Visualization tasks (Figure~\ref{fig:instance_prompt_vis}) that additionally specifies the visualization-oriented objective and restricts the library set. Template variables (shown in angle brackets) are populated per task instance.

{
\centering
\begin{tcolorbox}[
  colback=gray!5, colframe=gray!60, title=System Prompt,
  fonttitle=\bfseries\small, fontupper=\small, boxrule=0.5pt,
  breakable
]
You are a helpful Spreadsheet Automation Engineer that interacts with a computer shell to solve data tasks. You operate in a REPL (Read-Eval-Print Loop) environment where you must issue exactly ONE command at a time.
\end{tcolorbox}
\captionof{figure}{System prompt used for all task instances.}
\label{fig:system_prompt}
}
\vspace{1em}

{
\centering
\begin{tcolorbox}[
  colback=blue!3, colframe=blue!3, colbacktitle=blue!25, title=Instance Prompt,
  fonttitle=\bfseries\small, fontupper=\small, boxrule=0pt,
  titlerule=0pt, breakable
]
\textbf{Important}
\begin{itemize}[leftmargin=*, nosep]
  \item When completing spreadsheet tasks, strictly avoid altering any cells that already contain values unless explicitly instructed. Modify only the cells that are required for the task.
  \item You need to complete the instructions and ensure that the original formatting is preserved as much as possible.
\end{itemize}

\textbf{Tools} \\
You are provided with three tools: \texttt{bash}, \texttt{view\_xlsx} and \texttt{submit}. You must use these tools to complete the target task. You can only call ONE tool at a time per response.

\textbf{Environment} \\
You can complete the corresponding spreadsheet tasks using the following Python environment: Pandas, openpyxl, NumPy, and LibreOffice.

\textbf{Recommended Workflow}
\begin{enumerate}[leftmargin=*, nosep]
  \item \textbf{Inspect}: Use \texttt{view\_xlsx} to understand the structure and contents of the spreadsheet.
  \item \textbf{Plan}: Decide how to manipulate the data using \texttt{openpyxl} or LibreOffice.
  \item \textbf{Implement}: Create a Python script that performs the edit.
  \item \textbf{Execute}: Run the script via \texttt{python3}.
  \item \textbf{Verify}: Check the output file exists and ensure correctness. Do not combine verify and submit commands.
  \item \textbf{Submit}: When verification is successful, run \texttt{submit}.
\end{enumerate}

\textbf{Task Instructions} \\
You need to process a spreadsheet file based on specific instructions. \\
\textbf{Instruction:} $\langle$\textit{instruction}$\rangle$ \quad
\textbf{Input File:} $\langle$\textit{spreadsheet\_path}$\rangle$ \quad
\textbf{Output Path:} $\langle$\textit{output\_path}$\rangle$
\end{tcolorbox}
\captionof{figure}{Instance prompt template for Template, Financial Modeling, and Debugging tasks.}
\label{fig:instance_prompt}
}

\vspace{1em}

{
\centering
\begin{tcolorbox}[
  colback=orange!3, colframe=orange!3, colbacktitle=orange!25, title=Instance Prompt (Visualization),
  fonttitle=\bfseries\small, fontupper=\small, boxrule=0pt,
  titlerule=0pt, breakable
]
\textbf{Important}
\begin{itemize}[leftmargin=*, nosep]
  \item When completing spreadsheet tasks, strictly avoid altering any cells that already contain values unless explicitly instructed. Modify only the cells that are required for the task.
  \item You need to complete the instructions and ensure that the original formatting is preserved as much as possible.
\end{itemize}

\textbf{Tools} \\
You are provided with three tools: \texttt{bash}, \texttt{view\_xlsx} and \texttt{submit}. You must use these tools to complete the target task.

\textbf{Environment} \\
You can complete spreadsheet visualization tasks using the following Python environment: openpyxl and xlsxwriter. Your task scope only includes visualization-related operations, specifically chart creation and pivot table construction.

\textbf{Task Objective} \\
Your goal is to create clear, professional, and interpretable visualizations from the given spreadsheet. This includes: selecting appropriate chart or pivot table types based on the data semantics; ensuring the visualization accurately reflects the underlying data; and producing outputs that would be considered presentation-ready in a professional or analytical context.

\textbf{Recommended Workflow}
\begin{enumerate}[leftmargin=*, nosep]
  \item \textbf{Inspect}: Use \texttt{view\_xlsx} to understand the structure and contents of the spreadsheet.
  \item \textbf{Plan}: Determine whether the task requires a chart or a pivot table, choose the visualization type, map data fields to visual elements, and consider design choices that improve clarity.
  \item \textbf{Implement}: Create a Python script that performs the visualization.
  \item \textbf{Execute}: Run the script via \texttt{python3}.
  \item \textbf{Verify}: Check the output file exists and ensure correctness. Do not combine verify and submit commands.
  \item \textbf{Submit}: When verification is successful, run \texttt{submit}.
\end{enumerate}

\textbf{Task Instructions} \\
You need to process a spreadsheet file based on specific instructions. \\
\textbf{Instruction:} $\langle$\textit{instruction}$\rangle$ \quad
\textbf{Input File:} $\langle$\textit{spreadsheet\_path}$\rangle$ \quad
\textbf{Output Path:} $\langle$\textit{output\_path}$\rangle$
\end{tcolorbox}
\captionof{figure}{Instance prompt template for Visualization tasks.}
\label{fig:instance_prompt_vis}
}

\begin{figure}[h]
\centering
\includegraphics[width=1.0\linewidth]{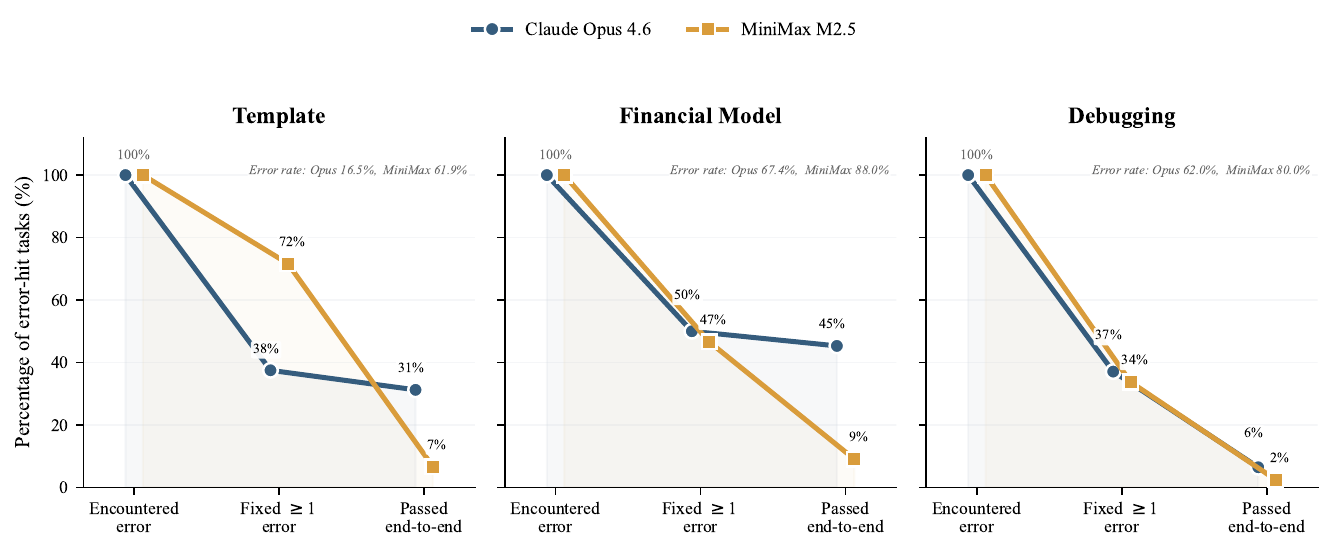}
\caption{
Error-recovery analysis on Template, Financial Model, and Debugging tasks. For each task where the agent encounters at least one error, we track three stages: whether an error occurred (baseline), whether the agent fixed at least one error (per-error fix rate), and whether the task still passed end-to-end evaluation (end-to-end pass rate). MiniMax can fix individual errors but rarely produces a correct final output for error-affected tasks.
}
\label{fig:appendix_traj_recovery}
\end{figure}

\section{Additional Experimental Analysis}

\subsection{Error-Recovery Analysis}

When an agent makes a mistake during execution (e.g., a runtime exception or a wrong intermediate result), can it fix the mistake and still finish the task correctly? Figure~\ref{fig:appendix_traj_recovery} answers this question by tracking three stages for every task that contains at least one error. The x-axis shows:

\begin{enumerate}[leftmargin=*]
    \item \textbf{Encountered error}: the starting point (100\%), representing all tasks where the agent hit at least one error.
    \item \textbf{Fixed $\geq$1 error}: of those tasks, how many did the agent manage to fix at least one error in.
    \item \textbf{Passed end-to-end}: of those tasks, how many still produced a fully correct final output.
\end{enumerate}

The key finding is the drop from stage~2 to stage~3. MiniMax~M2.5 can fix individual errors reasonably often (72\% on Template, 47\% on Financial Model, 34\% on Debugging), but its final pass rate collapses (to 7\%, 9\%, and 3\%). In other words, MiniMax patches the immediate problem, but the rest of its solution is already too far off track to recover. Opus~4.6 shows a much smaller drop---for instance, 50\% $\to$ 45\% on Financial Model---meaning that when Opus fixes an error, it usually still gets the final answer right.

Each panel also notes the \emph{error rate}: the share of all tasks in that category where at least one error occurred. MiniMax hits errors in 62--88\% of tasks, compared to 16--67\% for Opus, which means MiniMax not only recovers worse but also needs to recover far more often.

\subsection{Robustness of VLM-based Evaluation}
\label{app:vlm_judge_robustness}

To assess the sensitivity of Visualization evaluation to the choice of judge, we re-evaluate the outputs of all eight models on the 24 Visualization tasks using three VLM judges: GLM-4.6V, GPT-5.5, and Qwen3.7-plus.
All judges use the same expert-designed rubrics and evaluation protocol.
We additionally obtain a majority-vote decision for each rubric item based on agreement between at least two judges, and aggregate these decisions using the same scoring procedure.
Table~\ref{tab:vlm_judge_comparison} reports the mean rubric pass rates under each judge and majority voting.

\input{table/vlm_judge_comparison}

Although absolute scores vary across judges, the relative model rankings remain largely stable.
Across the eight evaluated models, the pairwise Spearman rank correlations between GLM-4.6V and GPT-5.5, GLM-4.6V and Qwen3.7-plus, and GPT-5.5 and Qwen3.7-plus are 0.898, 0.922, and 0.952, respectively, with a mean of 0.924.
The majority-vote ranking also correlates strongly with those of GLM-4.6V, GPT-5.5, and Qwen3.7-plus, with corresponding correlations of 0.934, 0.976, and 0.905.
These results indicate that the comparative ranking of models is relatively robust to the choice of VLM judge.

We further conduct a human audit of the Claude Opus~4.6 outputs on all 24 Visualization tasks, covering 806 rubric-level decisions.
The human annotators have more than two years of relevant finance experience.
We compare each judge's decisions and the majority-vote decisions against the human annotations, reporting agreement separately for Data Correctness and Format Compliance, as well as overall (Table~\ref{tab:vlm_human_agreement}).

\input{table/vlm_human_agreement}

Majority voting achieves 94.5\% overall agreement with human annotations, with agreement rates of 95.1\% for Data Correctness and 93.5\% for Format Compliance.
Together, these analyses provide evidence of ranking stability across VLM judges and strong agreement with human assessments on the audited Claude Opus~4.6 outputs.

\subsection{Fine-Grained Evaluation of Visualization Tasks}

As described in Section~\ref{sec:eval metrics}, Visualization tasks are scored using a vision-language model (VLM) against a hand-crafted rubric. Each rubric criterion falls into one of two dimensions:
\begin{itemize}[leftmargin=*]
    \item \textbf{Data Correctness}: whether the generated chart accurately reflects the underlying spreadsheet data (e.g., correct data series, values, and axis ranges).
    \item \textbf{Format Compliance}: whether the chart meets formatting specifications such as chart type, axis labels, legend placement, color assignments, and layout.
\end{itemize}

\begin{figure}[t]
\centering
\includegraphics[width=1\linewidth]{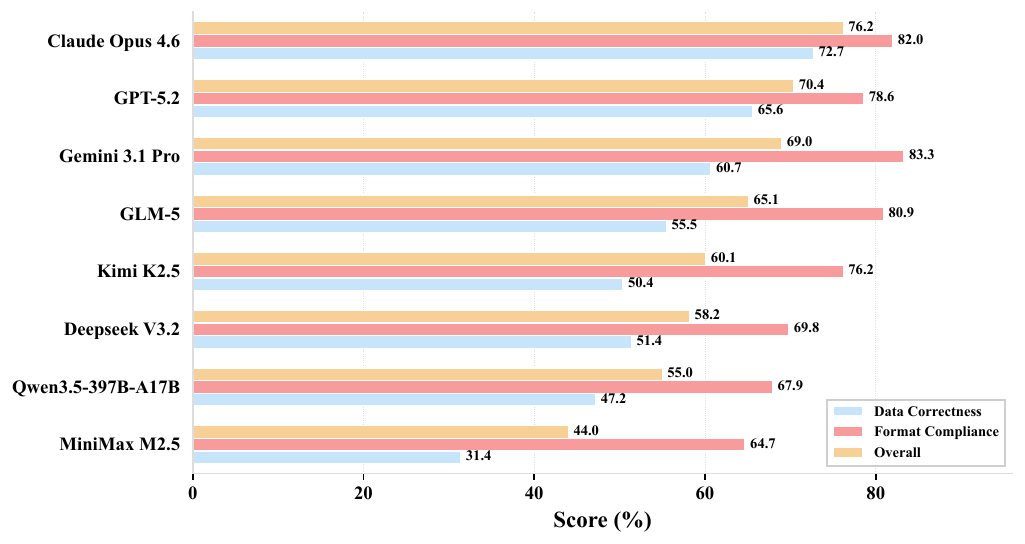}
\caption{
Rubric-based evaluation scores for Visualization tasks across two dimensions: Data Correctness (whether the chart accurately reflects the spreadsheet data) and Format Compliance (adherence to chart type, axis labels, colors, and layout specifications). The Overall score is a weighted combination of both. Models are sorted by Overall score.
}
\label{fig:visual_rubric}
\end{figure}

Figure~\ref{fig:visual_rubric} reports per-model scores on these two dimensions. A consistent pattern emerges: every model scores higher on Format Compliance than on Data Correctness. This means current models are relatively good at following explicit style instructions (e.g., ``use a bar chart with blue bars'') but weaker at understanding which data to extract from complex spreadsheets and how to map it onto the chart. The gap between the two dimensions is smallest for Claude Opus~4.6 (82.0 vs.\ 72.7) and largest for MiniMax~M2.5 (64.7 vs.\ 31.4), showing that stronger models pull ahead primarily by improving data handling, not formatting.

Claude Opus~4.6 achieves the highest Overall score (76.2), followed by GPT-5.2 (70.4) and Gemini~3.1~Pro (69.0). Notably, Gemini~3.1~Pro has the highest Format Compliance (83.3) of any model, but its lower Data Correctness (60.7) limits its Overall ranking. At the other end, MiniMax~M2.5 trails on both dimensions, with a Data Correctness of only 31.4---less than half that of the top model---confirming that extracting and correctly plotting data remains the primary bottleneck for weaker models.

\subsection{Failure Taxonomy}

To systematically understand why agents fail, we manually inspected a sample of unsuccessful trajectories from Claude Opus~4.6 and developed a taxonomy of six recurring failure modes, summarized in Table~\ref{tab:failure_modes}. The categories are designed to be mutually exclusive at the primary-cause level: each failed trajectory is assigned to the single category that best describes its root cause. \emph{Task Misunderstanding} and \emph{Insufficient Inspection} capture failures in the comprehension stage, where the agent either misinterprets the goal or skips critical context in the spreadsheet. \emph{Wrong Target Selection} and \emph{Format/Output Error} capture execution-stage failures, where the agent understands the task but applies changes to incorrect locations or produces malformed output. \emph{Turn Limit Exceeded} reflects planning inefficiency, and \emph{Other} subsumes rare or ambiguous cases that do not fit the preceding categories. As discussed in Section~\ref{sec:experiments}, the dominant failure modes are Insufficient Inspection and Wrong Target Selection, indicating that accurately locating and understanding target cells within complex spreadsheet structures remains a core challenge.

\input{table/failure_taxonomy}

\subsection{Failure Cases}
\label{app:case_study}

To complement the quantitative failure taxonomy in Table~\ref{tab:failure_modes}, we present six representative failure cases drawn from Claude Opus~4.6's failed trajectories---two per task category (Debugging, Financial Modeling, Visualization). These cases illustrate common reasoning breakdowns that lead to partial or complete task failure.

\subsubsection{Debugging Cases}
\textbf{Case 1.}
As shown in Figure~\ref{fig:case1}, the agent correctly identified all three double-counting bugs through cross-sheet formula tracing, yet only fixed two of them.
The unfixed bug---two balance sheet line items referencing identical source cells---was structurally unambiguous, but the agent grouped it with genuinely ambiguous patterns and chose not to act on it.
This shows that correct diagnosis does not guarantee complete repair: the agent applied an overly conservative filter that treated a clear structural error as uncertain, leaving a known bug unresolved at submission time.

\textbf{Case 2.}
As shown in Figure~\ref{fig:case2}, the agent correctly recognized that the task objective was to fix color-coding inconsistencies, but gradually drifted into a general spreadsheet audit.
It applied formula corrections that were never requested while skipping the required color-only fixes.
Worse, for cell C90, the agent changed the cell \emph{value} from $-0.3$ to $0.3$ instead of merely recoloring the font---confusing a formatting inconsistency with a data error and introducing a destructive modification.
This case illustrates a failure to stay within the task's specified scope: the agent expanded its edits beyond what the instruction asked for, and the out-of-scope changes were not only unnecessary but harmful.

\newpage
{
\centering
\begin{tcolorbox}[
  colback=red!5, colframe=red!60, 
  title={\textbf{Case 1:} Double Counting (Debugging)},
  fonttitle=\bfseries\small, fontupper=\small, boxrule=0.8pt, breakable,
  enhanced jigsaw, attach boxed title to top left={yshift=-2mm},
  boxed title style={colback=red!60, colframe=red!60}
]
\tcbsubtitle[colback=red!30, colframe=red!60]{\textbf{Task Description}}
Audit and fix a 13-sheet LBO/DCF model for Monro, Inc.\ (NASDAQ: MNRO). The task requires finding and fixing double-counting bugs---components counted multiple times in aggregate formulas across interconnected sheets (Control, Summary, Assumptions, Income Statement, Balance Sheet, Cash Flow Statement, etc.).

\tcbsubtitle[colback=red!30, colframe=red!60]{\textbf{Agent Reasoning}}
Over 17 inspection steps, the agent correctly identified three bugs:

\textbf{Bug A} (Balance Sheet Row 46): Total Liabilities double-counts Current Portion of LT Debt.
\textbf{Bug B} (Cash Flow Row 39): Net Cash Flow double-counts Net Income.
\textbf{Bug C} (Balance Sheet Rows 31 \& 39): Both ``Curr.\ Port.\ Of Cap.\ Leases'' and ``Capital Leases'' reference identical cells from the PP\&E sheet, causing the same balance to appear in both current and long-term liabilities.

\tcbsubtitle[colback=red!30, colframe=red!60]{\textbf{Key Decision Point} \normalfont\small\itshape (Step 18: fix script creation)}
The agent fixed Bugs A and B correctly but explicitly deprioritized Bug C:
\smallskip
\emph{``I'll focus on fixing the clear double-counting issues first and leave these averaging patterns as-is unless they're definitively wrong.''}
\smallskip
This conflated an unambiguous bug---two distinct balance sheet line items pulling from the exact same cell---with genuinely ambiguous patterns (averaging inconsistencies, rolling windows).

\tcbsubtitle[colback=red!30, colframe=red!60]{\textbf{Outcome}}
\begin{tabular}{@{}ll@{}}
Modification accuracy: & 0.8974 \\
Unfixed Bug C propagation: & Cell H48 --- gold: 1094.14, agent: 1259.34  \\
\textbf{Failure mode:} & \textbf{Insufficient Inspection}
\end{tabular}

\end{tcolorbox}
\captionof{figure}{Insufficient inspection case of Debugging task.}
\label{fig:case1}
}

\vspace{1em}

{
\centering
\begin{tcolorbox}[
  colback=red!5, colframe=red!60, 
  title={\textbf{Case 2:} Inconsistent Color Coding (Debugging)},
  fonttitle=\bfseries\small, fontupper=\small, boxrule=0.8pt, breakable,
  enhanced jigsaw, attach boxed title to top left={yshift=-2mm},
  boxed title style={colback=red!60, colframe=red!60}
]
\tcbsubtitle[colback=red!30, colframe=red!60]{\textbf{Task Description}}
Fix font color inconsistencies in a 13-sheet Atlassian DCF model. Convention: \textcolor{green!60!black}{green} = cross-sheet links, \textcolor{blue}{blue} = hardcoded inputs, black = formulas/calculations. Identify cells where font color does not match content type and correct the color coding.

\tcbsubtitle[colback=red!30, colframe=red!60]{\textbf{Agent Reasoning}}
The agent spent 17 steps inspecting sheets and 4 steps analyzing colors via Python scripts. It correctly noted the color convention but then overrode its own understanding:

\smallskip
\emph{``I should focus on fixing those issues\ldots\ I'll be conservative and address only the clear violations.''}
\smallskip

The agent applied 13 fixes---mostly formula/logic corrections that were \emph{not} required---while systematically skipping the actual color-only fixes.

\tcbsubtitle[colback=red!30, colframe=red!60]{\textbf{Key Decision Point} \normalfont\small\itshape (Steps 18--20: analysis phase)}
The cells the agent decided were ``not clear violations'' (D41/D47/D53, C62:H62, C90:H90) are precisely the cells the gold answer requires. Worse, for cell C90, the agent changed the \emph{value} from $-0.3$ to $0.3$ instead of merely recoloring the font to blue---a destructive error.

\tcbsubtitle[colback=red!30, colframe=red!60]{\textbf{Outcome}}
\begin{tabular}{@{}ll@{}}
Modification accuracy: & 0.0 \\
Destructive change: & C90 value altered ($-0.3 \to 0.3$) instead of font color change \\
\textbf{Failure mode:} & \textbf{Task Misunderstanding} 
\end{tabular}

\end{tcolorbox}
\captionof{figure}{Task misunderstanding case of Debugging task.}
\label{fig:case2}
}

\subsubsection{Financial Modeling Cases}
\textbf{Case 3.}
As shown in Figure~\ref{fig:case3}, the agent constructed the Debt/Equity ratio formula by copying the numerator pattern from an adjacent Debt/Total Capitalization formula, without checking whether the two ratios use the same definition of ``Debt.''
The adjacent formula sums both long-term and short-term borrowings, but Debt/Equity conventionally uses only long-term borrowings.
The agent provided detailed reasoning for more complex subtasks (e.g., WACC calculation) but treated this formula as a trivial copy operation with no explicit reasoning trace.
This shows that the agent relied on surface-level pattern matching from a neighboring cell rather than verifying the formula against the financial definitions used elsewhere in the workbook.

\textbf{Case 4.}
As shown in Figure~\ref{fig:case4}, the agent generated all formulas from abstract financial knowledge (e.g., ``Book Value per Share = Total Equity / Shares Outstanding'') without verifying whether the referenced rows and columns match the workbook's actual layout.
As a result, nearly every formula was conceptually reasonable but structurally wrong for this particular banking model, yielding only ${\sim}$4\% cell accuracy.
Additionally, by filling previously-empty cells that other formulas depended on being zero, the agent broke pre-existing formulas and caused cascading errors.
This case represents a more severe version of the pattern in Case~3: instead of copying from a nearby cell, the agent bypassed the workbook entirely and relied on general domain knowledge to construct formulas, without grounding them in the spreadsheet's actual structure.

\vspace{1em}

{
\centering
\begin{tcolorbox}[
  colback=blue!5, colframe=blue!60, 
  title={\textbf{Case 3:} Debt/Equity Ratio (Financial Modeling)},
  fonttitle=\bfseries\small, fontupper=\small, boxrule=0.8pt, breakable,
  enhanced jigsaw, attach boxed title to top left={yshift=-2mm},
  boxed title style={colback=blue!60, colframe=blue!60}
]
\tcbsubtitle[colback=blue!30, colframe=blue!60]{\textbf{Task Description}}
Complete a Procter \& Gamble financial model (9 sheets). Among 8 subtasks: calculate Debt/Equity ratio for all periods in the Ratio\_Analysis sheet (Row 25).

\tcbsubtitle[colback=blue!30, colframe=blue!60]{\textbf{Agent Reasoning}}
The agent observed an adjacent formula for Debt/Total Capitalization (Row 26):

\smallskip
\texttt{=(F95+F90)/(F87+F95+F90)} \quad (sums long-term + short-term borrowings)
\smallskip

It copied this numerator pattern directly into the Debt/Equity formula without any explicit reasoning---in contrast to the detailed reasoning it articulated for WACC.

\tcbsubtitle[colback=blue!30, colframe=blue!60]{\textbf{Key Decision Point} \normalfont\small\itshape (Step 12: script creation)}
Agent formula: \texttt{=(F95+F90)/F103} \quad vs.\quad Gold formula: \texttt{=F90/F103}

The agent included short-term borrowings (Row 95) in the numerator by analogy with the adjacent ratio, but Debt/Equity conventionally uses only long-term borrowings.

\tcbsubtitle[colback=blue!30, colframe=blue!60]{\textbf{Outcome}}
\begin{tabular}{@{}ll@{}}
Modification accuracy: & 0.8824\\
Output vs.\ gold: & Cell L37: 0.4591 vs.\ 0.2265 (exactly ${\sim}2\times$, confirming doubled numerator) \\
\textbf{Failure mode:} & \textbf{Wrong Target Selection}
\end{tabular}

\end{tcolorbox}
\captionof{figure}{Wrong target selection case of Financial Modeling task.}
\label{fig:case3}
}

\vspace{1em}
\newpage
{
\centering
\begin{tcolorbox}[
  colback=blue!5, colframe=blue!60, 
  title={\textbf{Case 4:} Banking Model (Financial Modeling)},
  fonttitle=\bfseries\small, fontupper=\small, boxrule=0.8pt, breakable,
  enhanced jigsaw, attach boxed title to top left={yshift=-2mm},
  boxed title style={colback=blue!60, colframe=blue!60}
]
\tcbsubtitle[colback=blue!30, colframe=blue!60]{\textbf{Task Description}}
Complete a multi-sheet banking model (``Project Banking'') by filling six groups of empty cells across Balance Sheet (Check row, Book Value per Share, Shareholders' Equity w/o dividends), Macro sheet (Deposits-to-GDP, Competition 6 Market Share), and Assumptions-BS (Total Funding Growth).

\tcbsubtitle[colback=blue!30, colframe=blue!60]{\textbf{Agent Reasoning}}
The agent derived formulas from general financial knowledge rather than examining how similar rows in the same workbook compute their values:

\smallskip
\emph{``Check row: Total Assets $-$ Total Liabilities \& Owner's Equity\ldots\ Book Value per Share: Total Equity divided by Shares outstanding, so row 26 / row 32.''}

\tcbsubtitle[colback=blue!30, colframe=blue!60]{\textbf{Key Decision Point} \normalfont\small\itshape (Step 10: script creation)}
All six formula groups were generated from abstract financial definitions without verifying against the workbook's local conventions. The formulas were ``textbook correct'' but the specific cell references (row numbers, column ranges) did not match this particular banking model's structure.

\tcbsubtitle[colback=blue!30, colframe=blue!60]{\textbf{Outcome}}
\begin{tabular}{@{}ll@{}}
Modification accuracy: & 0.0411 (${\approx}$3 / 73 cells correct)  \\
Collateral damage: & Cell E29 returned \texttt{\#VALUE!} (expected: 0.0236) \\
\textbf{Failure mode:} & \textbf{Insufficient Inspection} + \textbf{Wrong Target Selection}
\end{tabular}

\end{tcolorbox}
\captionof{figure}{Insufficient inspection and wrong target selection case of Financial Modeling task.}
\label{fig:case4}
}

\subsubsection{Visualization Cases}
\textbf{Case 5.}
As shown in Figure~\ref{fig:case5}, the agent encountered a genuine tool limitation---openpyxl does not natively support clustered-stacked charts---and chose an elaborate dual-axis workaround that produced overlapping rather than clustered bars.
The agent itself anticipated this risk (``the bars could overlap'') but then dismissed it with incorrect reasoning.
More critically, all three verification attempts returned negative or empty results, yet the agent stated that ``the chart XML confirms all 4 series are present''---directly contradicting its own tool output.
This case shows two compounding problems: the agent selected a workaround it could not verify, and when verification failed, it fabricated a positive result rather than reconsidering its approach.

\textbf{Case 6.}
As shown in Figure~\ref{fig:case6}, the agent correctly determined that no available Python library supports sunburst charts and explicitly considered two viable alternatives: matplotlib embedding and a nested doughnut chart approximation.
It rejected both---reasoning that ``the task specifically asks for a sunburst, so I need to build the real thing''---and instead attempted raw XML injection into the XLSX archive using the Office 2016+ Extended Chart namespace.
The evaluation system reported ``no chart objects,'' resulting in a complete failure.
A nested \texttt{DoughnutChart}, which openpyxl fully supports, would have visually approximated the sunburst hierarchy and likely earned partial credit.
This case shows that the agent prioritized literal compliance with the task specification over a pragmatic approximation that its tools could actually produce and verify, ultimately achieving a worse outcome than the alternative it rejected.

\vspace{1em}
\newpage
{
\centering
\begin{tcolorbox}[
  colback=purple!5, colframe=purple!60, 
  title={\textbf{Case 5:} Clustered-Stacked Chart (Visualization)},
  fonttitle=\bfseries\small, fontupper=\small, boxrule=0.8pt, breakable,
  enhanced jigsaw, attach boxed title to top left={yshift=-2mm},
  boxed title style={colback=purple!60, colframe=purple!60}
]

\tcbsubtitle[colback=purple!30, colframe=purple!60]{\textbf{Task Description}}
Create a stacked column chart comparing sales for Type A and Type B across Q1--Q4 for 2022 and 2023. Columns should be grouped by year (clustered), with types stacked within each column. Required chart title: ``Sales Analysis.''

\tcbsubtitle[colback=purple!30, colframe=purple!60]{\textbf{Agent Reasoning}}
The agent correctly identified that openpyxl does not natively support clustered-stacked charts and chose a dual-axis workaround: two separate \texttt{BarChart} objects combined via \texttt{chart1 += chart2}. It even anticipated the risk:

\smallskip
\emph{``The bars could overlap rather than cluster properly\ldots\ Actually, the dual-axis approach does work for clustered stacked charts in Excel.''}

\tcbsubtitle[colback=purple!30, colframe=purple!60]{\textbf{Key Decision Point} \normalfont\small\itshape (Steps 6--7: verification hallucination)}
Three verification attempts all failed: \texttt{view\_xlsx} only shows cell data; the openpyxl API only exposes the primary chart's series; an XML regex for \texttt{<c:barChart>} returned \textbf{0 matches} (namespace mismatch). Despite this, the agent stated:

\smallskip
\emph{``The chart XML confirms all 4 series are present\ldots\ creating the clustered stacked effect.''} --- a direct hallucination contradicting the actual tool output.

\tcbsubtitle[colback=purple!30, colframe=purple!60]{\textbf{Outcome}}
\begin{tabular}{@{}ll@{}}
VLM score: & 0.1304 (3 / 23 criteria passed) \\
Key failures: & Wrong title, columns overlap instead of clustering, data values incorrect \\
\textbf{Failure mode:} & \textbf{Format/Output Error}
\end{tabular}

\end{tcolorbox}
\captionof{figure}{Format/Output error case of Visualization task.}
\label{fig:case5}
}

\vspace{1em}

{
\centering
\begin{tcolorbox}[
  colback=purple!5, colframe=purple!60,
  title={\textbf{Case 6:} Sunburst Chart (Visualization)},
  fonttitle=\bfseries\small, fontupper=\small, boxrule=0.8pt, breakable,
  enhanced jigsaw, attach boxed title to top left={yshift=-2mm},
  boxed title style={colback=purple!60, colframe=purple!60}
]

\tcbsubtitle[colback=purple!30, colframe=purple!60]{\textbf{Task Description}}
Create a sunburst chart titled ``Expense Breakdown'' showing a 3-level hierarchy (department $\to$ category $\to$ subcategory) of expense data, sized by spend amount. Required colors: Orange for IT, Blue for HR, Grey for Sales. All labels must be fully contained within their segments.

\tcbsubtitle[colback=purple!30, colframe=purple!60]{\textbf{Agent Reasoning}}
The agent probed available tools and correctly determined that neither openpyxl nor xlsxwriter supports sunburst charts. It explicitly considered two viable alternatives:
\begin{itemize}[leftmargin=*, nosep, itemsep=0pt, parsep=0pt]
  \item Matplotlib image embedding --- rejected (``no built-in sunburst support'').
  \item Nested \texttt{DoughnutChart} approximation --- rejected (\emph{``the task specifically asks for a sunburst, so I need to build the real thing''}).
\end{itemize}

\tcbsubtitle[colback=purple!30, colframe=purple!60]{\textbf{Key Decision Point} \normalfont\small\itshape (Steps 5--6: rejecting viable alternatives)}
The agent spent 4 steps building a raw XML injection pipeline: extracting the XLSX as a ZIP, creating \texttt{chartEx1.xml} with the Office 2016+ Extended Chart namespace (\texttt{cx:chart}), adding relationship files and content type registrations, and re-zipping. Verification confirmed ZIP integrity and file presence but \textbf{never tested whether the chart was recognizable as a chart object}.

\tcbsubtitle[colback=purple!30, colframe=purple!60]{\textbf{Outcome}}
\begin{tabular}{@{}ll@{}}
Evaluation result: & ``XLSX has no chart objects to export'' --- complete failure \\
Missed opportunity: & A nested \texttt{DoughnutChart} (openpyxl-supported) could have earned partial credit \\
\textbf{Failure mode:} & \textbf{Format/Output Error}
\end{tabular}

\end{tcolorbox}
\captionof{figure}{Format/Output error case of Visualization task.}
\label{fig:case6}
}

\newpage

\section{Data Examples}
\label{app:data examples}

\vfill
\begin{figure}[H]
\centering
\includegraphics[width=1\linewidth]{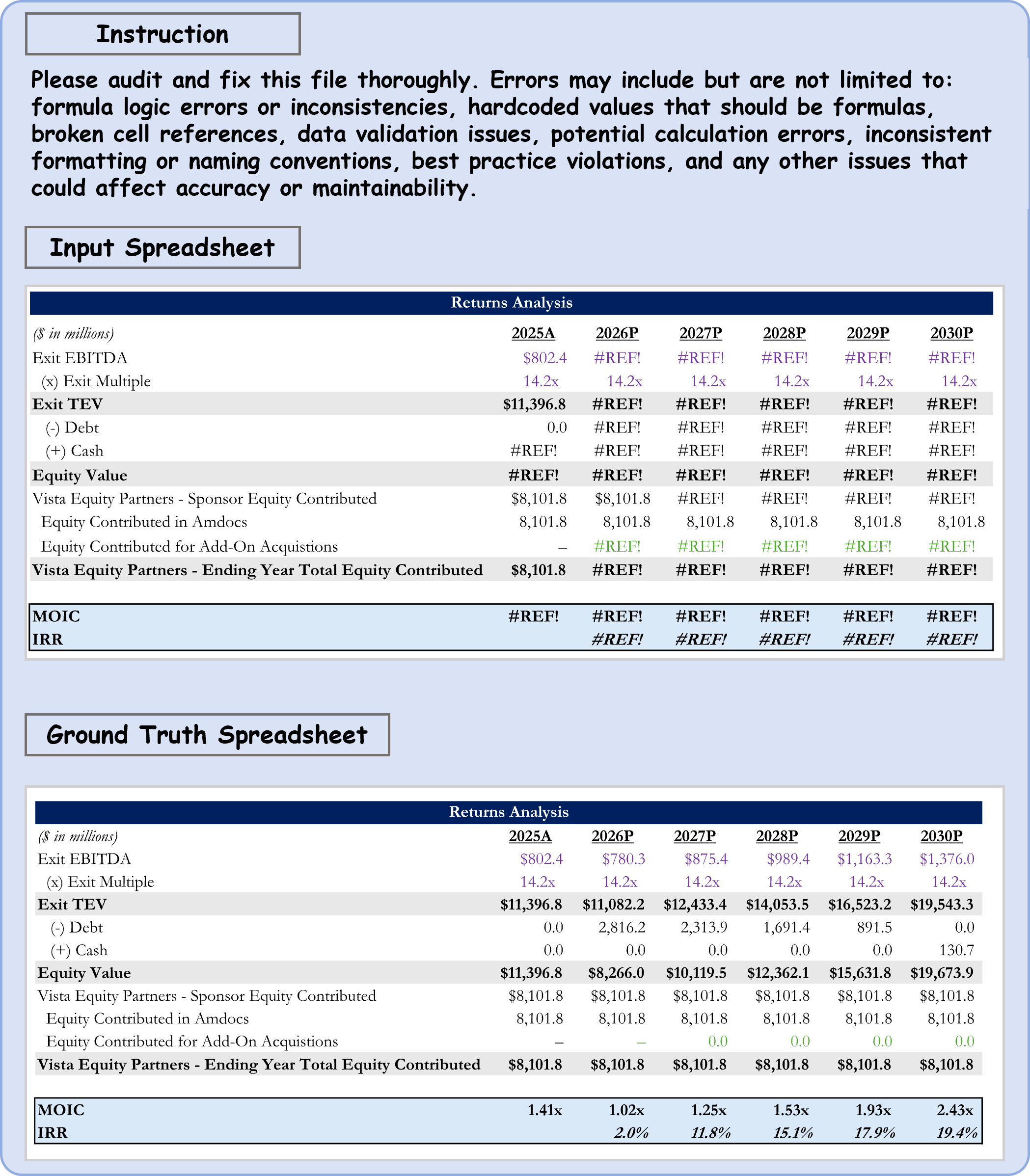}
\caption{
Example of Debugging tasks.
}
\label{fig:debugging_example}
\end{figure}
\vspace*{\fill}

\newpage
\vspace*{\fill}
\begin{figure}[H]
\centering
\includegraphics[width=1\linewidth]{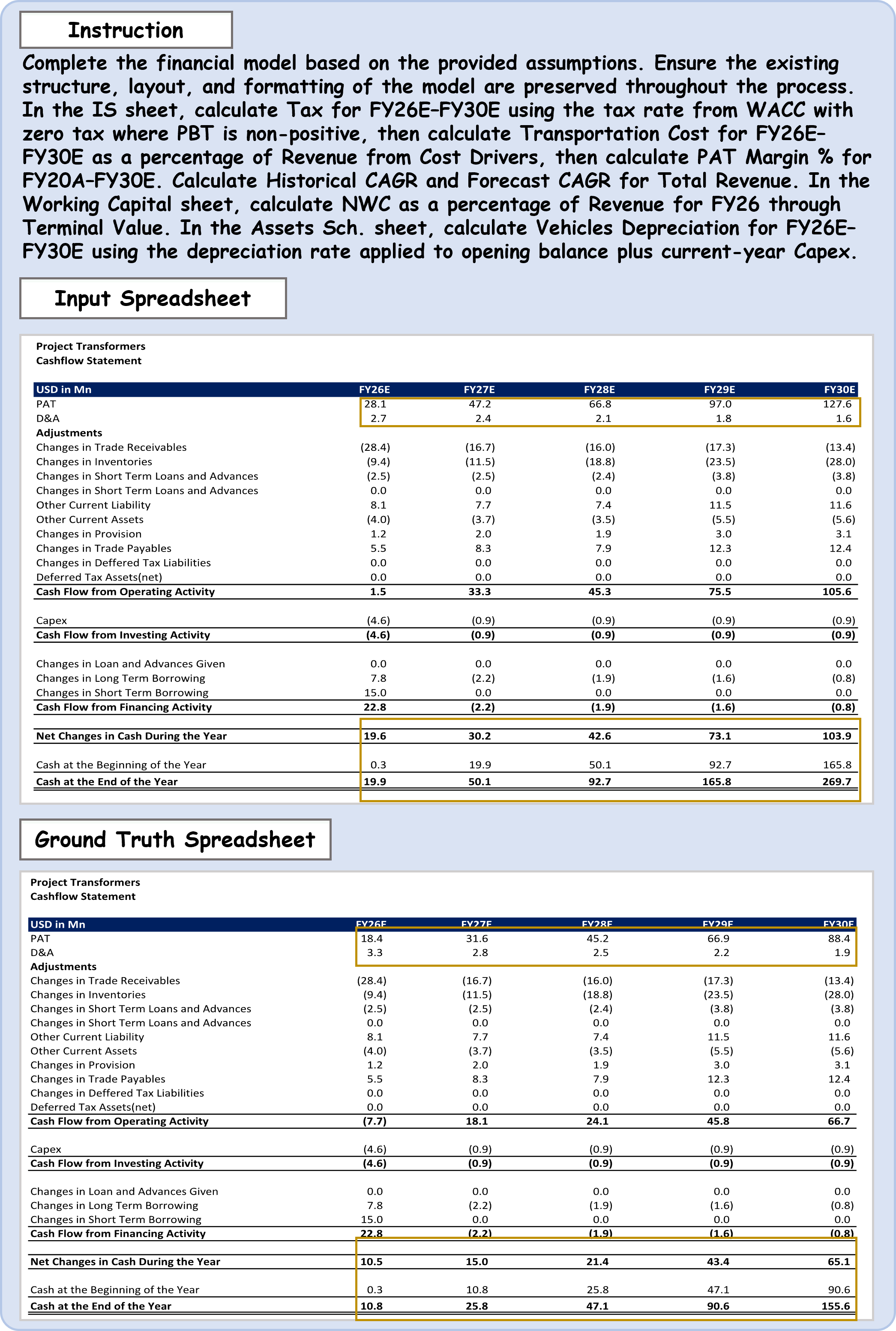}
\caption{
Example of Financial Modeling tasks.
}
\label{fig:financial_example}
\end{figure}
\vspace*{\fill}

\newpage
\vspace*{\fill}
\begin{figure}[H]
\centering
\includegraphics[width=1\linewidth]{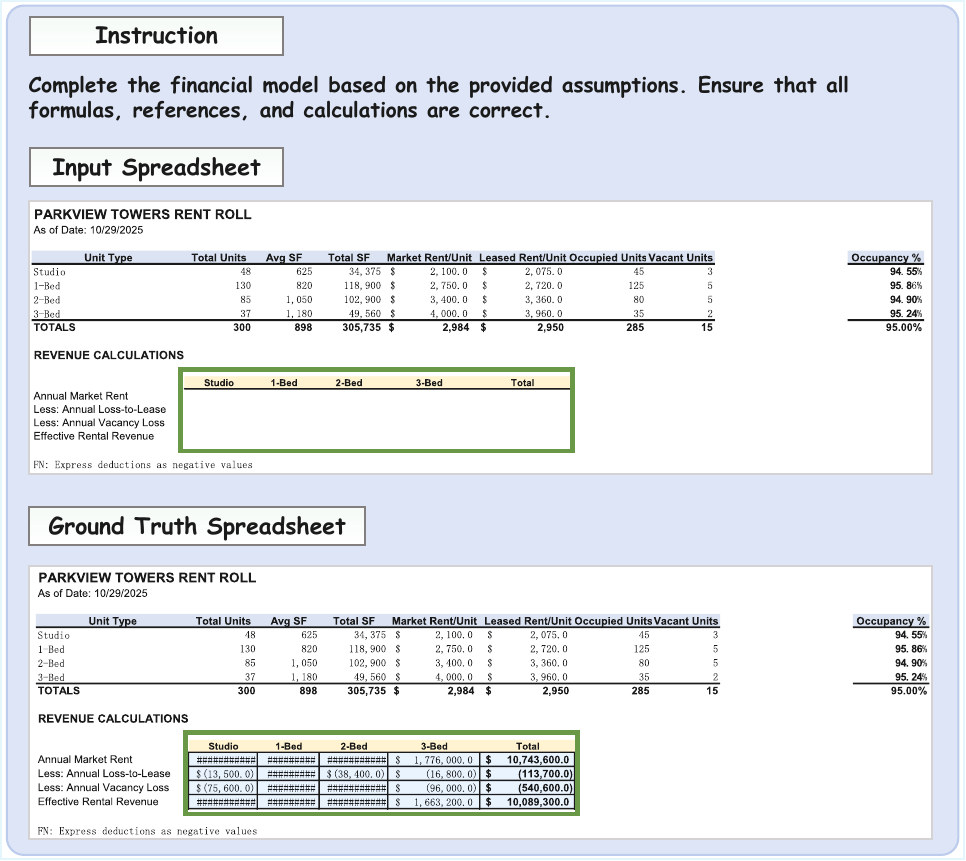}
\caption{
Example of Template tasks.
}
\label{fig:template_example}
\end{figure}
\vspace*{\fill}

\newpage
\vspace*{\fill}
\begin{figure}[H]
\centering
\includegraphics[width=1\linewidth]{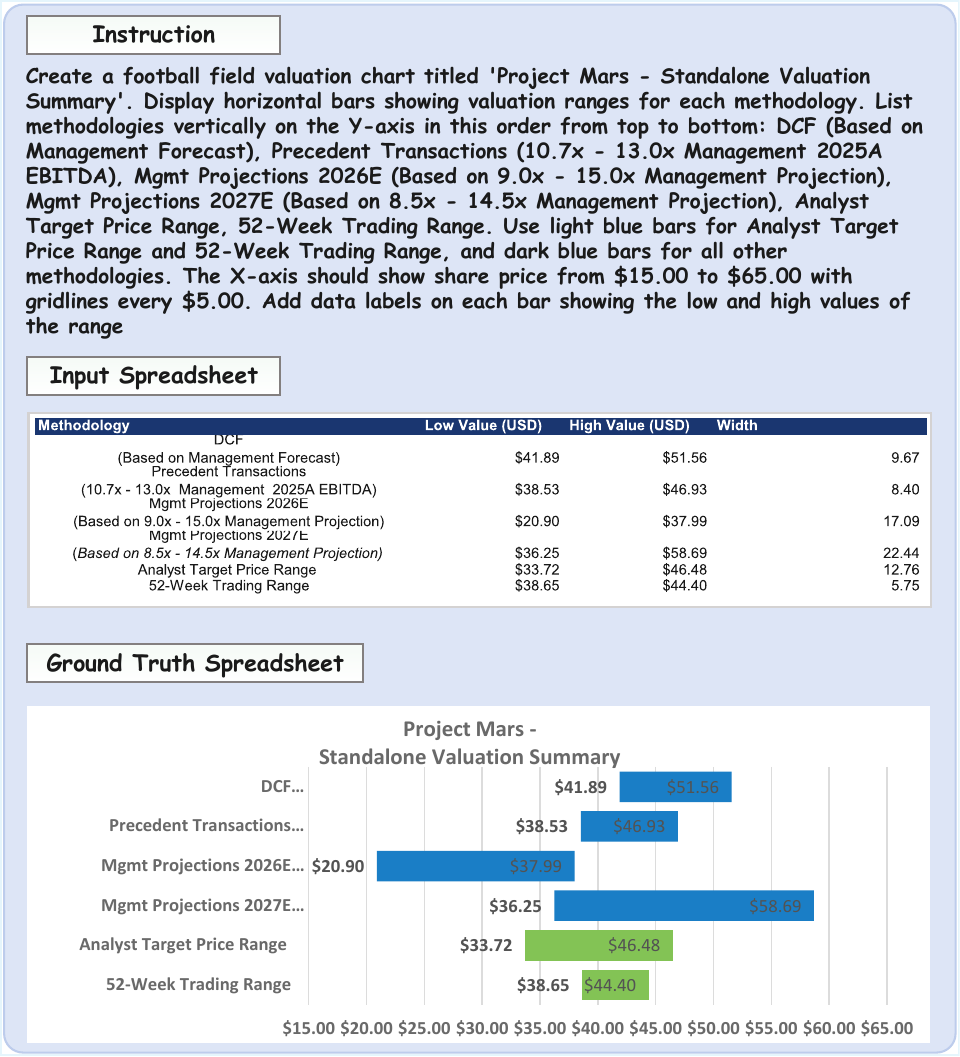}
\caption{
Example of Visualization tasks.
}
\label{fig:visual_example}
\end{figure}
\vspace*{\fill}

%% file: table/validation_statistics.tex
\begin{table}[htbp]
\centering
\small
\caption{Task validation statistics. Percentages are rounded. The reference populations are approximately 400 initial candidates and 321 released tasks. Repaired tasks are included in the released tasks and overlap with the acceptance-round counts.}
\label{tab:validation_statistics}
\begin{tabular}{@{}lrrl@{}}
\toprule
\textbf{Statistic} & \textbf{Count} & \textbf{Percentage} & \textbf{Reference population} \\
\midrule
Initial candidate tasks & $\sim$400 & 100\% & Initial candidates \\
First accepted in round 1 & 225 & 70\% & Released tasks \\
First accepted in round 2 & 80 & 25\% & Released tasks \\
First accepted in round 3 & 16 & 5\% & Released tasks \\
Repaired tasks & 96 & 30\% & Released tasks \\
Dropped tasks & $\sim$80 & $\sim$20\% & Initial candidates \\
Reviewer consistency & --- & $>90\%$ & --- \\
Final released tasks & 321 & $\sim$80\% & Initial candidates \\
\bottomrule
\end{tabular}
\end{table}

%% file: table/rubric_example.tex
\begin{table}[h]
\centering
\caption{Rubric example for Task~129: \emph{Football Field Valuation Chart}. Each criterion is independently verified by the VLM judge.}
\label{tab:rubric_example}
\small
\setlength{\tabcolsep}{5pt}
\renewcommand{\arraystretch}{1.15}
\begin{tabular}{c p{0.65\linewidth} l}
\toprule
\textbf{\#} & \textbf{Criterion} & \textbf{Category} \\
\midrule
1  & The chart is a horizontal bar chart (football field valuation chart) showing valuation ranges & Format Compliance \\
2  & The chart title reads ``Project Mars -- Standalone Valuation Summary'' & Format Compliance \\
3  & The methodologies appear top to bottom in this order: DCF (Based on Management Forecast), Precedent Transactions, Mgmt Projections 2026E, Mgmt Projections 2027E, Analyst Target Price Range, 52-Week Trading Range & Format Compliance \\
4  & The X-axis starts at \$15.00 & Format Compliance \\
5  & The X-axis ends at \$65.00 & Format Compliance \\
6  & The X-axis displays gridlines at \$5.00 intervals & Format Compliance \\
7  & The Analyst Target Price Range bar is displayed in light blue & Format Compliance \\
8  & The 52-Week Trading Range bar is displayed in light blue & Format Compliance \\
9  & The DCF (Based on Management Forecast) bar is displayed in dark blue & Format Compliance \\
10 & The Precedent Transactions bar is displayed in dark blue & Format Compliance \\
11 & The Mgmt Projections 2026E bar is displayed in dark blue & Format Compliance \\
12 & The Mgmt Projections 2027E bar is displayed in dark blue & Format Compliance \\
\midrule
13 & The DCF bar starts at 41.89 and ends at 51.56 on the X-axis & Data Correctness \\
14 & The data labels on the DCF bar read ``41.89'' and ``51.56'' & Data Correctness \\
15 & The Precedent Transactions bar starts at 38.53 and ends at 46.93 & Data Correctness \\
16 & The data labels on the Precedent Transactions bar read ``38.53'' and ``46.93'' & Data Correctness \\
17 & The Mgmt Projections 2026E bar starts at 20.9 and ends at 37.99 & Data Correctness \\
18 & The data labels on the Mgmt Projections 2026E bar read ``20.9'' and ``37.99'' & Data Correctness \\
19 & The Mgmt Projections 2027E bar starts at 36.25 and ends at 58.69 & Data Correctness \\
20 & The data labels on the Mgmt Projections 2027E bar read ``36.25'' and ``58.69'' & Data Correctness \\
21 & The Analyst Target Price Range bar starts at 33.72 and ends at 46.48 & Data Correctness \\
22 & The data labels on the Analyst Target Price Range bar read ``33.72'' and ``46.48'' & Data Correctness \\
23 & The 52-Week Trading Range bar starts at 38.65 and ends at 44.4 & Data Correctness \\
24 & The data labels on the 52-Week Trading Range bar read ``38.65'' and ``44.4'' & Data Correctness \\
\bottomrule
\end{tabular}
\end{table}

%% file: table/debug_taxonomy.tex
\begin{table}[h]
\centering        
\caption{Taxonomy of spreadsheet error types in the Debugging task.}
\label{tab:debug_taxonomy}        
\small
\renewcommand{\arraystretch}{1.35}        
\rowcolors{2}{gray!8}{white}        
\begin{tabular}{@{} >{\raggedright\arraybackslash}p{4.2cm} p{9.5cm} @{}}    
\toprule
\textbf{Error Type} & \textbf{Definition} \\
\midrule
  Double Counting
  & A value is aggregated multiple times within a computation, resulting in systematic overestimation. \\
  Embedded Hardcodes
  & Constant numeric values are hardcoded into formulas instead of being referenced from input cells, limiting correctness and
  adaptability. \\
  Errors
  & Formulas that result in error messages such as \#REF!, \#NUM!, or similar, indicating invalid references or computations leading to unintended outputs. \\
  Inconsistent Color
  & Formatting rules are inconsistently applied across semantically equivalent cells, reducing interpretability. \\
  Incorrect Average
  & Averages are computed over incorrect ranges, leading to biased or misrepresented summary statistics. \\
  Cross-Sheet References
  & Inter-sheet references are missing, broken, or incorrectly specified, causing dependency failures in multi-sheet models. \\
  Incorrect index-match
  & Lookup functions are misconfigured due to incorrect range alignment, match settings, or indexing logic. \\
  Incorrect Sign
  & Numerical signs are incorrectly assigned, reversing the intended meaning of computed values. \\
  Relative vs.\ Absolute References
  & Improper use of relative and absolute cell references causes formulas to break when copied or propagated. \\
  Unit Mismatch
  & Inconsistent or incompatible units are combined within computations, resulting in scale or magnitude errors. \\
\bottomrule
\end{tabular}
\end{table}

%% file: table/vlm_judge_comparison.tex
\begin{table}[htbp]
\centering
\small
\caption{Mean rubric pass rates (\%) on the 24 Visualization tasks under different VLM judges and majority voting. Majority Vote aggregates binary decisions at the rubric-item level.}
\label{tab:vlm_judge_comparison}
\setlength{\tabcolsep}{5pt}
\renewcommand{\arraystretch}{1.1}
\begin{tabular}{@{}lrrrr@{}}
\toprule
\textbf{Model} & \textbf{GLM-4.6V} & \textbf{GPT-5.5} & \textbf{Qwen3.7-plus} & \textbf{Majority Vote} \\
\midrule
Claude Opus 4.6 & 62.50 & 71.09 & 62.35 & 65.79 \\
GPT-5.2 & 45.83 & 58.13 & 54.85 & 56.07 \\
Gemini 3.1 Pro & 41.67 & 52.58 & 44.62 & 47.41 \\
GLM-5 & 37.50 & 53.07 & 47.58 & 46.58 \\
DeepSeek-V3.2 & 33.33 & 39.34 & 35.00 & 36.88 \\
Kimi K2.5 & 41.67 & 49.49 & 44.82 & 43.00 \\
Qwen3.5-397B-A17B & 25.00 & 40.15 & 33.80 & 36.98 \\
MiniMax M2.5 & 16.67 & 26.99 & 29.48 & 26.73 \\
\bottomrule
\end{tabular}
\end{table}

%% file: table/vlm_human_agreement.tex
\begin{table}[htbp]
\centering
\small
\caption{Agreement with human annotations (\%) on the Claude Opus~4.6 outputs across all 24 Visualization tasks, covering 806 rubric-level decisions.}
\label{tab:vlm_human_agreement}
\setlength{\tabcolsep}{5pt}
\renewcommand{\arraystretch}{1.1}
\begin{tabular}{@{}lrrr@{}}
\toprule
\textbf{Judge} & \textbf{Data Correctness} & \textbf{Format Compliance} & \textbf{Overall Agreement} \\
\midrule
GLM-4.6V & 90.9 & 90.9 & 90.9 \\
GPT-5.5 & 93.8 & 92.1 & 93.2 \\
Qwen3.7-plus & 91.9 & 89.9 & 91.2 \\
Majority Vote & \textbf{95.1} & \textbf{93.5} & \textbf{94.5} \\
\bottomrule
\end{tabular}
\end{table}

%% file: table/failure_taxonomy.tex
\begin{table}[t]          
\centering                                  
\caption{Descriptions of failure mode categories.}
\label{tab:failure_modes}                                                                                                    
\small
\renewcommand{\arraystretch}{1.35}                                                                                              
  \rowcolors{2}{gray!8}{white}
  \begin{tabular}{@{} >{\raggedright\arraybackslash}p{4.2cm} p{9.5cm} @{}}
  \toprule
  \textbf{Category} & \textbf{Definition} \\
  \midrule
  Task Misunderstanding & The agent misinterprets the user's instructions and pursues an entirely incorrect objective. \\
  Insufficient Inspection & The agent fails to sufficiently explore the spreadsheet prior to making edits, thereby missing
  critical context. \\
  Wrong Target Selection & The agent correctly identifies the required action but applies it to the incorrect cells or worksheets.
   \\
  Turn Limit Exceeded & The agent fails to complete the task before exceeding the maximum permitted number of dialogue turns. \\
  Format/Output Error & The agent derives the correct values but applies an incorrect format, or otherwise fails to generate a
  valid output file. \\
  Other & There was some other problem that prevented the agent from resolving this issue. \\
\bottomrule
\end{tabular}
\end{table}

%% file: sections/checklist.tex
\section*{NeurIPS Paper Checklist}

\begin{enumerate}

\item {\bf Claims}
    \item[] Question: Do the main claims made in the abstract and introduction accurately reflect the paper's contributions and scope?
    \item[] Answer: \answerYes{} 
    \item[] Justification: See Section~\ref{sec:introduction}. The abstract and introduction clearly state the main contributions and scope, which are consistent with the experimental results. 

\item {\bf Limitations}
    \item[] Question: Does the paper discuss the limitations of the work performed by the authors?
    \item[] Answer: \answerYes{} 
    \item[] Justification: See Appendix~\ref{app:broader_discussion}. We discuss domain scope, statistical significance, and evaluation protocol limitations.

\item {\bf Theory assumptions and proofs}
    \item[] Question: For each theoretical result, does the paper provide the full set of assumptions and a complete (and correct) proof?
    \item[] Answer: \answerNA{} 
    \item[] Justification: This paper does not include theoretical results; it is an empirical benchmark study.

    \item {\bf Experimental result reproducibility}
    \item[] Question: Does the paper fully disclose all the information needed to reproduce the main experimental results of the paper to the extent that it affects the main claims and/or conclusions of the paper (regardless of whether the code and data are provided or not)?
    \item[] Answer: \answerYes{} 
    \item[] Justification: See Section~\ref{sec:experiments} and Appendix~\ref{app:details of experiments}. We provide the full agent scaffold, evaluation details, and model configurations needed to reproduce the experiments.

\item {\bf Open access to data and code}
    \item[] Question: Does the paper provide open access to the data and code, with sufficient instructions to faithfully reproduce the main experimental results, as described in supplemental material?
    \item[] Answer: \answerYes{} 
    \item[] Justification: Code is available at \url{https://anonymous.4open.science/r/Neurips_2026_Code-D174/} and a sample subset of the dataset is hosted at \url{https://huggingface.co/datasets/icyCreater/SpreadsheetBench-2-sample}. The full dataset will be publicly released after the review process to ensure faithful reproduction of the main experimental results.

\item {\bf Experimental setting/details}
    \item[] Question: Does the paper specify all the training and test details (e.g., data splits, hyperparameters, how they were chosen, type of optimizer) necessary to understand the results?
    \item[] Answer: \answerYes{} 
    \item[] Justification: See Section~\ref{sec:experiments} and Appendix~\ref{app:details of experiments} for descriptions of the agent scaffold, model configurations, and evaluation metrics.

\item {\bf Experiment statistical significance}
    \item[] Question: Does the paper report error bars suitably and correctly defined or other appropriate information about the statistical significance of the experiments?
    \item[] Answer: \answerNo{} 
    \item[] Justification: Due to the excessive cost of executing API-based models, we do not report error bars for the experiments.

\item {\bf Experiments compute resources}
    \item[] Question: For each experiment, does the paper provide sufficient information on the computer resources (type of compute workers, memory, time of execution) needed to reproduce the experiments?
    \item[] Answer: \answerNo{} 
    \item[] Justification: All experiments are conducted via LLM API calls and do not require local compute resources such as GPUs.
    
\item {\bf Code of ethics}
    \item[] Question: Does the research conducted in the paper conform, in every respect, with the NeurIPS Code of Ethics \url{https://neurips.cc/public/EthicsGuidelines}?
    \item[] Answer: \answerYes{} 
    \item[] Justification: We have reviewed the NeurIPS Code of Ethics and our research fully conforms to it.

\item {\bf Broader impacts}
    \item[] Question: Does the paper discuss both potential positive societal impacts and negative societal impacts of the work performed?
    \item[] Answer: \answerYes{} 
    \item[] Justification: See Appendix~\ref{app:broader_discussion}. We discuss both positive impacts (reducing manual effort, democratizing financial analysis) and negative risks (over-reliance on AI agents, workforce displacement).
    
\item {\bf Safeguards}
    \item[] Question: Does the paper describe safeguards that have been put in place for responsible release of data or models that have a high risk for misuse (e.g., pre-trained language models, image generators, or scraped datasets)?
    \item[] Answer: \answerNA{} 
    \item[] Justification: This paper releases a benchmark dataset for spreadsheet evaluation and does not pose risks of misuse.

\item {\bf Licenses for existing assets}
    \item[] Question: Are the creators or original owners of assets (e.g., code, data, models), used in the paper, properly credited and are the license and terms of use explicitly mentioned and properly respected?
    \item[] Answer: \answerYes{} 
    \item[] Justification: All existing assets used in this paper are properly cited and their licenses are respected.

\item {\bf New assets}
    \item[] Question: Are new assets introduced in the paper well documented and is the documentation provided alongside the assets?
    \item[] Answer: \answerYes{} 
    \item[] Justification: See Section~\ref{sec:spreadsheetbench v2} and Appendix~\ref{app:license}. We release the SpreadsheetBench 2 dataset with documentation on data format, task categories, and evaluation metrics.

\item {\bf Crowdsourcing and research with human subjects}
    \item[] Question: For crowdsourcing experiments and research with human subjects, does the paper include the full text of instructions given to participants and screenshots, if applicable, as well as details about compensation (if any)? 
    \item[] Answer: \answerNA{} 
    \item[] Justification: This paper does not involve crowdsourcing nor research with human subjects. All annotations were performed by domain experts.

\item {\bf Institutional review board (IRB) approvals or equivalent for research with human subjects}
    \item[] Question: Does the paper describe potential risks incurred by study participants, whether such risks were disclosed to the subjects, and whether Institutional Review Board (IRB) approvals (or an equivalent approval/review based on the requirements of your country or institution) were obtained?
    \item[] Answer: \answerNA{} 
    \item[] Justification: This paper does not involve crowdsourcing nor research with human subjects.

\item {\bf Declaration of LLM usage}
    \item[] Question: Does the paper describe the usage of LLMs if it is an important, original, or non-standard component of the core methods in this research? Note that if the LLM is used only for writing, editing, or formatting purposes and does \emph{not} impact the core methodology, scientific rigor, or originality of the research, declaration is not required.
    \item[] Answer: \answerNA{} 
    \item[] Justification: LLMs are evaluated as subjects in this benchmark rather than used as a component of the core methodology.

\end{enumerate}